\documentclass[a4paper,11pt]{article}
\usepackage{jcappub} 
\usepackage[T1]{fontenc} 
\usepackage{graphics}  
\usepackage{graphicx}                                                          
\usepackage{color}                                                              
\usepackage{ifthen}                                                             
\usepackage{mathrsfs}
\usepackage{amsmath, amssymb}
\usepackage{multirow}                                                          
\usepackage{tabularx}
\usepackage{titlesec}
\def\ds{{\rm d}s}
\def\da{{\rm d}a}

\def\d{{\rm d}}
\title{Morphology of CMB fields - effect of weak gravitational lensing}

\author[a,b]{Priya Goyal}
\author[a,1]{Pravabati Chingangbam}{\note{corresponding author}}
\author[c]{Stephen Appleby}

\affiliation[a]{Indian Institute of Astrophysics, Koramangala II Block,       
  Bangalore  560 034, India}
\affiliation[b]{Department of Physics, Pondicherry University, R.V. Nagar, Kalapet, 605014, Puducherry, India}
\affiliation[c]{Quantum Universe Center, Korea Institute for Advanced Study, 85 Hoegiro, Dongdaemun-gu,
  Seoul 02455, South Korea}  

\emailAdd{priya.goyal@iiap.res.in}
\emailAdd{prava@iiap.res.in}
\emailAdd{stephen@kias.re.kr}

\abstract{
  We study the morphology of the cosmic microwave background temperature and polarization fields using the shape and alignment parameters, $\beta$ and $\alpha$, that are constructed from the contour Minkowski tensor. The primary goal of our paper is to understand the effect of weak gravitational lensing on the morphology of the CMB fields. In order to isolate different physical effects that can be potentially confused with the effect of lensing, we first study the effect of varying the cosmology on $\alpha$ and $\beta$, and show that they are relatively insensitive to variation of cosmological parameters. Next we analyze the signatures of hemispherical anisotropy, and show that information of such anisotropy in $\alpha$ gets washed out at small angular scales and become pronounced only at large angular scales. For $\beta$ we find characteristic distortions which vary with the field threshold.  We then study the effect of weak gravitational lensing using simulations of lensed temperature and $E$ and $B$ modes.  We quantify the distortion induced in the fields across different angular scales. We find that lensing makes structures of all fields increasingly more anisotropic as we probe down to smaller scales. We find distinct behaviour of morphological distortions as a function of threshold for the different fields. The effect is small for temperature and $E$ mode, while it is significantly large for $B$ mode. Further, we find that lensing does not induce statistical anisotropy, as expected from the isotropic distribution of large scale structure of matter. We expect that the results obtained in this work will provide insights on the reconstruction of the lensing potential.}

\begin{document}

\maketitle
\flushbottom

\section{Introduction} 
\label{sec:intro}


The statistical properties of the cosmic microwave backrground (CMB) radiation have revealed a wealth of information about the origin, physical properties and time evolution of the universe. 
Also encoded in the statistical properties is the information of how the CMB photons have been deflected multiple times  by gravitational potential wells that they encounter along their path from the last scattering surface to an observer~\cite{Blanchard:1987}.
The rms of the deflection angles in different sky directions can be estimated to be of the order of a few  arcminutes~\cite{Blanchard:1987,Lewis:2006fu}. While the blackbody frequency spectrum of the CMB remains unchanged, the cumulative effect of the path deflections of the photons translates into a remapping of the CMB total intensity and polarization components from one sky direction to another. The angular separation between the original and remapped directions depend on the distribution of matter and its time evolution that the photons encounter in their paths. It is therefore related to the integral along the lines of sight of the gravitational potential field.   
This remapping results in redistribution of power at different scales, and as a consequence it leads to smoothening of the peaks in the CMB angular power spectra~\cite{Seljak:1995ve,Metcalf:1997ih,Zaldarriaga:1998ar}. 
An important effect of lensing on the CMB is the generation of $B$-modes at small angular scales due to the coupling between $E$ and $B$-mode~\cite{Zaldarriaga:1998ar}. This means that gravitational lensing can generate $B$-mode polarization, even if no tensor primordial fluctuations were present to generate primordial $B$-mode polarization. 
Weak gravitational lensing of the CMB was first confirmed by Smith et. al.~\cite{Smith:2007rg} by cross-correlating CMB data from WMAP~\cite{Hinshaw:2006ia} and galaxy data~\cite{WISE}. It has now been confirmed by Planck at the level of $40\sigma$~\cite{Aghanim:2018oex}. $B$-mode sourced by lensing was first detected by the South Polar telescope~\cite{SPT:2013}, and subsequently by Planck~\cite{Ade:2015nch}.  These detections have opened up exciting new avenues to probe the universe. 


Most statistical analyses of CMB anisotropies are carried out using $n$-point functions in harmonic space, namely, the angular power spectrum, bispectrum, and so on. 
The observed CMB data agree very well with Gaussian statistics, at the precision level of currently available cosmological datasets. Therefore, the need for going beyond the 2-point function, such as in searches for non-Gaussian deviations, is served well by the next or next-to-next higher order statistics. In real space the analysis of the rich geometrical and topological properties of excursion sets of smooth random fields provides a suite of statistical observables that can provide information that is complementary to Fourier or harmonic analysis. In some cases, such as when dealing with strongly non-Gaussian fields, it can be more profitable to employ real space statistical tools that can, in principle, encode all orders of $n$-point functions. The reason is that in such cases one needs to go well beyond the 2-point functions to extract all statistical information, and calculating $n$-point functions to arbitrary order is not realistic for large cosmological datasets.

One important class of real space statistics are the so called  
scalar Minkowski Functionals (henceforth MFs) which are defined for excursion sets of smooth random fields. The definitions rely on a mathematical framework that is borne out of a interplay between integral geometry and probability theory~\cite{Adler:1981,Tomita:1986}. They have been widely used in  cosmology~\cite{Gott:1990,Mecke:1994,Schmalzing:1997,Schmalzing:1998}. 
They  have been  applied  to  study  primordial  non-Gaussianity  using  CMB  temperature  and $E$-mode fields~\cite{COBE_NG:2000,WMAP_NG:2011,Chingangbam:2009vi,Ganesan:2014lqa,Chingangbam:2017sap,Ade:2015ava,Buchert:2017uup}. They have also been applied to study CMB foregrounds~\cite{Chingangbam:2013,Rana:2018oft}. 
Non-Gaussianity of lensed $B$-mode has been investigated using MFs in ~\cite{Santos:2015}. Betti numbers, which are the numbers of connected regions and holes and whose difference give the genus (the third scalar Minkowski Functional), have also been  studied for cosmological fields~\cite{Chingangbam:2012,Park:2013,Pratyush:2018}. 
Vector- and tensor-valued generalizations of the scalar Minkowski Functionals on two- and three-dimensional Euclidean space have been constructed in~\cite{McMullen:1997,Alesker:1999, Hug:2008}. They are collectively known as tensor Minkowski functionals or Minkowski tensors. In comparison to scalar Minkowski Functionals they carry additional information related to intrinsic anisotropy and alignment of structures~\cite{Schroder2D:2009,Schroder3D:2013}. Of these,  rank-1 Minkowski tensors have been used to study the sub-structure of galaxy clusters~\cite{Beisbart:2001a,Beisbart:2001b}. Rank-2 Minkowski tensors can be further sub-classified into translation covariant and translation invariant ones. Translation covariant MTs have been used to study the sub-structure of spiral galaxies in~\cite{Beisbart:2002}.  

The {\em translation invariant rank-2 MTs} have been recently introduced to cosmological applications. They were first applied to analyze the CMB in \cite{Vidhya:2016}. The definition of MTs was generalized to smooth random fields on curved two-dimensional manifolds, in particular spaces of constant curvature such as the sphere, in~\cite{Chingangbam:2017uqv}.  The ensemble expectation values for isotropic Gaussian random fields on two-and three-dimensional spaces were derived in~\cite{Chingangbam:2017uqv, Appleby:2018tzk}. The numerical computation of these statistics on Euclidean space is described in~\cite{Vidhya:2016, Appleby:2018tzk,Appleby:2017uvb}, while corresponding calculation on the unit sphere is described in~\cite{Chingangbam:2017uqv}. They have also been applied to search for departure from statistical isotropy of the CMB~\cite{Vidhya:2016,Joby:2018}. They have also been applied to analyze fields of the epoch of reionization~\cite{Kapahtia:2017qrg,Kapahtia:2019ksk}. Most recently they have been used to study the effect of redshift space distortion on matter distribution in the universe~\cite{Appleby:2019nit}. 


Weak gravitational lensing causes shearing and magnification of the hotspots and coldspots of the CMB fields. By understanding the resulting distortion patterns one may then, in principle, infer the integrated effect of the  interaction between the CMB and the intervening matter distribution along each line of sight. The morphological information encoded in the Minkowski tensors make them well suited to probe the size and shape distortions of the structures of the CMB. In this paper we employ the contour Minkowski tensor, which is one of the translation invariant Minkowski tensors, to capture the distortions in the alignment and anisotropy induced by lensing on the structures in the CMB temperature and polarization fields. We quantify the distortion in the shape of hotspots and coldspots across the full range of the field, and determine the effect on the statistical isotropy. We find that structures in the lensed field are more anisotropic compared to the structures in unlensed field.
We carry out our analysis using unlensed and lensed maps simulated using the publicly available code \texttt{LENSPIX}~\cite{Lenspix}. 
In addition, we have analyzed the effect of varying cosmology, as well as a toy model of temperature field having hemispherical anisotropy, on morphological properties of the structures. We find that each of these cases have effects that are distinct from that of lensing and hence can be easily distinguished.

The results presented in this paper pave the way for further probes of the effect of lensing on the CMB and we expect that they will provide insights into the understanding and reconstruction of the lensing potential. It is important to mention that the applicability of our method relies on availability of observed data which are sufficiently high resolution, good coverage of large angular scales and high signal to noise ratio. These factors can limit their usefulness. 
Our work is similar in spirit to~\cite{Bernardeau:1998mw} where the author analyzed the distortion induced by weak lensing on the ellipticity of hotspots and coldspots of CMB temperature. The definition of the ellipticity parameter used in~\cite{Bernardeau:1998mw} was  derived from the Hessian matrix of the field. The lensing effect was quantified by comparing the probability distribution function (PDF) of ellipticity parameter of the lensed field and the unlensed field. It was found the lensing broadens the ellipticity PDF, which implies that the structures especially the peaks (or the extrema) in the field are elongated due to the lens effect.

This paper is organized as follows. In section~\ref{sec:sec2} we review the definition of the contour Minkowski Tensor, and the shape and alignment parameters obtained from it. In section~\ref{sec:sec3} we show $\alpha$ and $\beta$ for Gaussian isotropic maps of CMB temperature and $E$ and $B$ modes. We also discuss their sensitivity to variation of cosmological parameters  and  hemispherical anisotropic. 
In section~\ref{sec:sec4} we review the physics of CMB lensing and give a brief overview of simulation of lensed maps by {\texttt{LENSPIX}}. Then we present our results for unlensed and lensed CMB fields and discuss in detail the effect of lensing on the morphology of the fields. We end with a discussion of our results in section~\ref{sec:sec5}. Appendix~\ref{sec:appendA} discusses details of the effect of stereographic projection on the calculation of the anisotropy parameter.  
Details of the sensitivity of $\alpha$ and $\beta$ to cosmological parameters is given in appendix~\ref{sec:alpha_vary_lcdm} , while sensitivity to hemispherical anisotropy is given in appendix~\ref{sec:alpha_ha}.

\section{Contour Minkowski Tensor - definition, properties and computation}
\label{sec:sec2}

This section presents the definition of the contour Minkowski tensor, the statistics for measuring intrinsic anisotropy and alignment obtained from it, and our methods for numerical calculation.

\subsection{Definition}
The notation in this section follows~\cite{Chingangbam:2017uqv}. The contour Minkowski tensor (henceforth CMT) is defined for a smooth closed curve, $C$, on a general smooth two-dimensional manifold as,  
\begin{eqnarray}
  {\mathcal W}_1 &\equiv& \int_C \, \hat{T} \otimes \hat{T} \,\ds, 
\label{eqn:tmf_def}
\end{eqnarray}
where $\ds$ is the infinitesimal arc length, and $\hat T$ is the unit tangent vector at each point of the curve with the direction chosen to be one of the two possibilities. The tensor product $\otimes$ is defined to be symmetric and given by
\begin{equation}
\left(\hat{T} \otimes \hat{T}\right)_{ij} \equiv \frac12 \left( \hat{T}_i\hat{T}_j + \hat{T}_j\hat{T}_i \right).
\end{equation}
The trace of ${\mathcal W}_1 $ gives the perimeter of the curve which is the scalar contour Minkowski functional, $W_1$, 
\begin{eqnarray}
  {\mathbf {Tr}}\left({\mathcal{W}}_1\right) &=& \int_C \, \ds = W_1,
  \label{eqn:tmf_trace1} 
\end{eqnarray}
where we have used $|\hat T|^2=1$.  We can express  $\mathcal{W}_1$ as
\begin{equation}
  \mathcal{W}_1 = \left(
  \begin{array}{cc}
    \tau+g_1 & g_2\\
    g_2 & \tau-g_1
  \end{array}
\right),
\end{equation}
  where
\begin{eqnarray}
  \tau &=& \frac12 \int_C \,\ds,\\
  g_1 &=& \frac12 \int_C \, \left( \hat{T}_1^2-\hat{T}_2^2\right)\,\ds,\\
  g_2 &=& \int_C \, \hat{T}_1\hat{T}_2\,\ds.
\label{eqn:W1_tensor}
\end{eqnarray}
If we assume that the curve occupies a region that is small such that we can ignore the curvature of the manifold, then it is easy to see that under rotations the pair $(g_1,g_2)$ transforms as a rank-2 tensor, while $\tau$ and $g\equiv \sqrt{g_1^2+g_2^2}$ are scalars. Then the eigenvalues of $\mathcal{W}_1$ are given by
\begin{eqnarray}
  \lambda_1 &=& \tau-g,\\
  \lambda_2 &=& \tau+g.
\end{eqnarray}
Since $\mathcal{W}_1$ is real and symmetric the eigenvalues are real and  positive. Further, we can show $\tau \ge g$  and hence $\lambda_1<\lambda_2$.
\subsection{Shape and alignment parameters}
\label{sec:mtaniso}

The {\em shape anisotropy parameter} $\beta$ is defined to be the ratio of the eigenvalues,
\begin{equation}
  \beta\equiv\frac{\lambda_1}{\lambda_2}.
\end{equation}
$\beta$ lies between 0 and 1. It gives a measure of intrinsic anisotropy of the curve. It is equal to one for a closed curve having $m$-fold symmetry with $m\ge 3$. Deviation of  $\beta$ from one indicates anisotropy.

For a distribution of many curves, let us denote $\overline{\mathcal{W}}_1$ as the average of $\mathcal{W}_1$ over all structures, then we can define the {\em alignment parameter},
\begin{equation}
  \alpha\equiv\frac{\Lambda_1}{\Lambda_2},
\end{equation}
where $\Lambda_1$ and $\Lambda_2$ are the eigenvalues of $\overline{\mathcal{W}}_1$ such  that $\Lambda_1\leq\Lambda_2$. By definition we have $0\leq\alpha\leq1$. $\alpha$ gives a measure of the orientation or the deviation from rotational symmetry in the distribution of curves. For randomly oriented structures with no preferred direction $\alpha = 1$, else $\alpha$ lies between 0 and 1. For a single curve we have $\alpha=\beta$. For many curves $\alpha$ gives the $\beta$ value for the curve which is resultant from translating and stacking all the curves such that their centroids overlap.

\subsection{Contour Minkowski tensor for smooth random fields}
 \label{sec:w1_random}

 Let $f$ denote a random field on the sphere for which we wish to compute ${\cal W}_1$. We will work with the mean subtracted field, $u\equiv f-\mu$, where $\mu$ is the mean of $f$.  We then choose field threshold values denoted by $\nu$, and  then the set of all points on the manifold where the field has  values greater than or equal to $\nu$ form an excursion set, $\mathcal{Q}_{\nu}$. It consists of connected regions and holes. The boundaries, denoted by  $\partial\mathcal{Q}_{\nu}$, of these connected regions and holes form closed curves. 

Let $\nabla u = \left( u_{;1},u_{;2} \right)$ be the covariant derivative of $u$.  The components of the unit tangent vector can be expressed in terms of the field derivatives as,
 \begin{equation}
  \hat{T}_i = \epsilon_{ij} \, \frac{u_{;j}}{\left| \nabla u \right|},
  \label{eqn:hatt}
 \end{equation}
 where $\epsilon_{ij}$ is the Levi-Civita tensor in two dimensions. 
 Then ${\mathcal W}_1$ for the  $k^{th}$ curve in $\partial\mathcal{Q}_{\nu}$ can be expressed as,
 \begin{equation}
   {\mathcal W}_1(\nu,k) = \int_{C_k}  \ds  \,\frac{1}{|\nabla u|^2} \ {\mathcal M}, 
  \label{eqn:w1final}
 \end{equation}
where the matrix $\mathcal{M}$ is given by,
 \begin{equation}
  \mathcal M=  \left(
  \begin{array}{cc} 
    u_{;2}^2 &  u_{;1} \,u_{;2} \\
      u_{;1} \,u_{;2} & u_{;1}^2
  \end{array}\right).
  \label{eqn:M}
 \end{equation}
The sum of ${\mathcal W}_1$ (denoted with an overbar) over all curves in $\partial\mathcal{Q}_{\nu}$ is then
 \begin{equation}
   {\overline{\mathcal W}}_1 = \int_{\partial{\mathcal Q}_{\nu}}  \ds  \,\frac{1}{|\nabla u|^2} \ {\mathcal M}.
  \label{eqn:w1_all}
 \end{equation}
 Note that in cosmology the scalar Minkowski functionals are usually expressed as per unit area. By dividing the above equation by the total area of the manifold under consideration we can obtain ${\overline{\mathcal W}}_1$ per unit area.
 
\subsection{Numerical computation of the Contour Minkowski tensor}
 \label{sec:computemt}

We use two different methods for numerically computing $\mathcal{W}_1$. We briefly describe them in the following.

\subsubsection{Method 1 - using field derivatives}
\label{sec:method1}

The line integral over the boundary of $\mathcal{Q}_{\nu}$ in eq.~\ref{eqn:w1_all} can be transformed into an area integral by introducing a Jacobian to give
\begin{eqnarray}
   \overline{\mathcal {W}}_1 &=& \int_{{\mathcal S}^2} \da \, \, \delta(u-\nu)\ \frac{1}{|\nabla u|} \ {\mathcal M},
  \label{eqn:w1_area}
 \end{eqnarray}
where $\da$ is the infinitesimal area element on  ${\mathcal S}^2$ and $\delta(u-\nu)$ is the Dirac delta function. 
In order to numerically calculate Eq.~\eqref{eqn:w1_area}, the $\delta-$function can be approximated as (\cite{Schmalzing:1998,Chingangbam:2017uqv}),
 \begin{equation}
   \delta \left( u-\nu \right) = 
\left\{ \begin{array}{l}
    \frac{1}{\Delta \nu}, \quad {\rm if} \ 
   u \in \left( \nu-\frac{\Delta \nu}{2} , \nu+\frac{\Delta \nu}{2} \right)\\
    0, \quad {\rm otherwise},
  \end{array}
  \right.
  \label{eqn:delta}
 \end{equation}
 where $\Delta \nu$ is the bin size of the threshold values. 
 Eq.~\ref{eqn:w1_area} can be diagonalized, and the ratio of the eigenvalues give $\alpha$, by definition. {\em Note that we cannot obtain $\beta$ from this method since it does not isolate individual curves at each $\nu$.}

 When we work with finite resolution maps in a space of compact extent, such as the surface of a sphere,  even for a field which is given to be statistically isotropic, we do not obtain $\alpha$ to be exactly equal to one at each threshold  
 even for a field which is given to be statistically isotropic. $\alpha=1$ is recovered only in the limit of the total perimeter tending to infinity. At threshold values close to zero, where the perimeter of $\mathcal{Q}_{\nu}$ is the largest, $\alpha$ is closest to one. For a random distribution of a few curves, the probability that they will be isotropically distributed is  small.  As a result, at higher $|\nu|$, the values of $\alpha$ decrease from unity. It is, therefore, important to take into account the threshold dependence of $\alpha$ when searching for and interpreting results of statistical isotropy  on finite resolution fields. 

This method has inherent numerical error coming from the discrete approximation of the delta function~\cite{Lim:2012}. This numerical error has analytic form and is a function of the threshold bin width and the smoothing scale. As shown in~\cite{Chingangbam:2017uqv} the numerical error for the diagonal elements are similar in shape and amplitude for Gaussian isotropic fields (see figure~4 of the above reference). In order to make a rough estimate of the resulting error on $\alpha$, we express the diagonal elements of   ${\overline{\mathcal W}}_1$ as
 \begin{eqnarray}
   ({\overline{\mathcal W}}_1)_{11}(\nu) &=& a(\nu) + e(\nu)\nonumber\\
   ({\overline{\mathcal W}}_1)_{22}(\nu) &=& b(\nu) + e(\nu),
 \end{eqnarray}
 where $a(\nu)$ and $b(\nu)$ denote the true functional forms of the elements, and $e(\nu)$ denotes the numerical error coming from the approximate $\delta$ function. We have used $a$ and $b$ so as to keep the discussion general but they are expected to be comparable and differ at only very high threshold values (at several sigmas away from the mean value of the field), and exactly the same in the limit of high resolution. We have also used the same $e(\nu)$ for the two matrix elements which means that we are ignoring possible statistical fluctuations between them. Then $\alpha$ is given by
 \begin{equation}
\alpha(\nu) \simeq   \frac{({\overline{\mathcal W}}_1)_{11}(\nu)}{({\overline{\mathcal W}}_1)_{22}(\nu)} = \frac{a(\nu) + e(\nu)}{b(\nu) + e(\nu)} = \frac{a}{b} \left(1+\frac{e}{a}-\frac{e}{b}+\frac{e^2}{ab}+\mathcal{O}(e^3)\right),
 \end{equation}
 where on the right hand side we have used $|e/b| \ll 1$. We have also taken the off-diagonal element ${\overline{\mathcal W}}_{12}$ to be zero. This is because the ensemble expectation of the correlation between $u_{;1}$ and $u_{;2}$ is expected to be zero since they are independent random fields. Since $a\sim b$ the numerical error is  the numerical error term is $\mathcal{O}(e^2)$. Thus, this method of computation is well suited for applications where $\alpha$ is used to extract physical information.

\subsubsection{Method 2 - using identification of contours in pixel space}
\label{sec:method2}

An alternative method of numerically reconstructing  ${\cal W}_{1}$ was adopted in~\cite{Appleby:2017uvb} (see also~\cite{Vidhya:2016}). In this section we briefly review the approach. 

The continuous field $u$ is sampled on a uniformly spaced, two dimensional pixel grid to generate a discrete function $u_{ij}$, where $i,j$ denotes the $i^{\rm th}$, $j^{\rm th}$ pixel in the $x_{1},x_{2}$ coordinate space. For a given threshold $\nu$, we decompose the pixels as `inside' the excursion set if $u_{ij} > \nu$ and `outside' if $u_{ij} < \nu$. We then apply the marching squares algorithm and linearly interpolate between any two adjacent `in' and `out' pixels to create a set of vertices at which $u=\nu$. Finally these vertices are joined by line segments to create a boundary of constant threshold $u = \nu$. From the vertices and corresponding line segments that define the excursion set boundary, all topological quantities of interest can be deduced, including the scalar and tensor Minkowski functionals.

In this discretized setup, $\overline{\cal W}_{1}$ for 
the excursion set boundary ${\partial\cal Q}_{\nu}$ is approximated as 
\begin{equation} ({\overline{\cal W}}_{1})_{ij} =  \int_{\partial{\cal Q}_{\nu}} \hat{e}_{i} \hat{e}_{j} \ds =  \sum_{e} |\vec{e}|^{-1} e_{i}e_{j} \end{equation} 

\noindent where $\sum_{e}$ is the sum over all edges in the discretized boundary reconstruction and $e_{i}$, $|\vec{e}|$ are the length in the $i^{\rm th}$ direction, and total length, of an edge segment.

To construct $\beta$ from a field we must calculate ${\cal W}_{1}$ for each individual connected component within the excursion set, and also for every hole. To do so, we must not only assign each pixel $u_{ij}$ as `in' or `out' of the excursion set, but also a unique identifier that informs which distinct connected region or hole the pixel belongs to. This latter condition is achieved by using a simplified `friends-of-friends' algorithm, assigning an initial pixel to an excursion set, then all adjacent pixels which satisfy $u_{i'j'}>\nu$. This is repeated until every `in' pixel is assigned to an excursion set region. The procedure is then repeated for holes -- regions outside the excursion set. Each edge segment to the excursion set boundary $\vec{e}$ is associated with exactly one connected component and one hole -- ${\cal W}_{1}$ is calculated for each sub-region as the sum over edge segments that define its boundary.  

The principle source of error in using this method arises from the assumption that the field can be linearly interpolated between adjacent pixels. Critical points of a field are intrinsically higher order quantities, and hence the method will fail to accurately represent structures of size $\sim {\cal O}(\Delta^{2})$, where $\Delta$ is the resolution of the pixel grid. This issue is ameliorated by smoothing the field with scale $R_{\rm G}$, and it was shown in~\cite{Appleby:2017uvb} that smoothing over five pixel lengths $R_{\rm G} > 5 \Delta$ is sufficient to reduce the numerical error on the ${\cal W}_{1}$ statistic to below $1\%$ for threshold values in the range $-4 < \nu < 4$. 

For calculating $\beta$ for the CMB fields we first project each hemisphere of the CMB maps onto a plane and then implement the above algorithm. We choose stereographic projection, as done in~\cite{Vidhya:2016}, because it is a conformal mapping which preserves angles and shape of structures. However, it is important to note that sizes of structures are not preserved. The distortion is largest towards the outer edge of the hemisphere. Moreover, the distortion decreases as we probe smaller structures. A discussion of the effects of the stereographic projection on the structures can be found in~\cite{Vidhya:2016}. In appendix~\ref{sec:appendA} we have included details of the effect of pixel resolutions, on the sphere as well as the plane, on the accuracy of $\beta$ calculation. 
In the subsequent sections we do not attempt to correct for the numerical error arising from  this projection. We proceed to use this method of calculating $\beta$ with the anticipation that the error will cancel out when we compare $\beta$ for fields with a fixed realization seed but with varying physical input.  We will comment on this point further when we discuss specific results. 

\section{$\alpha$ and $\beta$ for CMB fields}
\label{sec:sec3}

The threshold dependence of $\alpha$ and $\beta$ for CMB temperature and $E$ mode were previously studied in~\cite{Vidhya:2016,Joby:2018}. Here we include $B$ mode in our analysis and discuss the broad features of $\alpha$ and $\beta$ for the three fields for the standard $\Lambda$CDM cosmology with assumed Gaussianity and isotropy. We also discuss the dependence of their functional forms on the smoothing scale. In subsequent sections we will use the symbol $T$ to refer to CMB temperature maps. 
For the calculation of $\alpha$ in this section and in section~\ref{sec:sec4} we use method 1 described in section~\ref{sec:method1}.
For calculating $\beta$ we use the method 2 described in section~\ref{sec:method2}. All maps in this section are made using \texttt{HEALPIX}~\cite{Gorski:2005,Healpix} and \texttt{CAMB}~\cite{Lewis:2000ah,cambsite}.

\subsection{$\alpha$ and $\beta$ for Gaussian isotropic $\Lambda$CDM cosmology}
\label{sec:alpha_lcdm}

\begin{figure}
  \includegraphics[height=3.cm,width=5.cm]{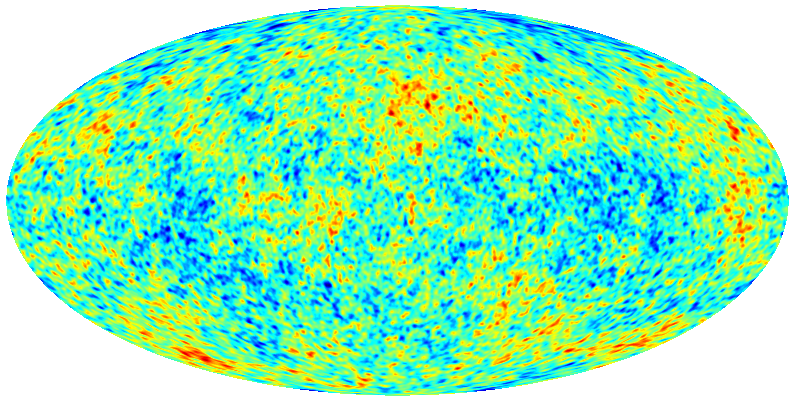}
  \includegraphics[height=3.cm,width=5.cm]{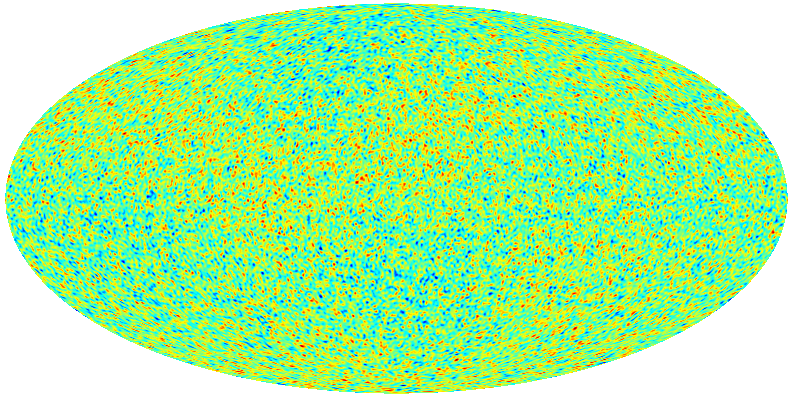}
  \includegraphics[height=3.cm,width=5.cm]{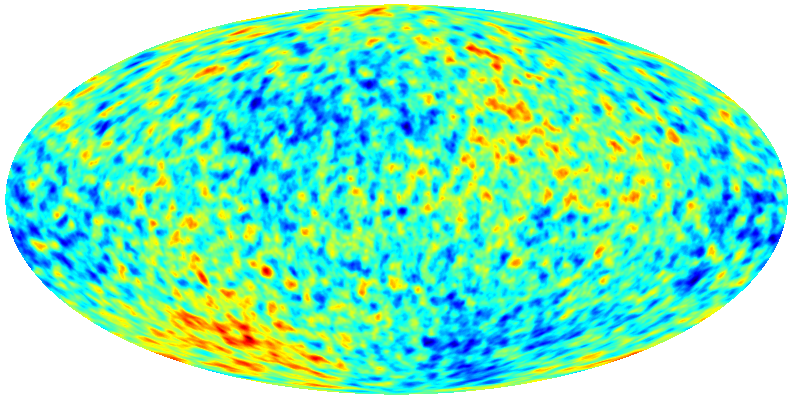}  
\caption{Maps of $T$ (left), $E$ (middle), and $B$ (right) with tensor to scalar ratio $r=0.1$. All maps are smoothed with FWHM=$1^{\circ}$. We have not put the colour scale since we wish to only draw visual attention to the fact that $E$ mode has the highest number of structures per unit area, while $B$ mode has the lowest. As a consequence $\alpha$ is expected to be largest for $E$ mode and lowest for $B$ mode, at all threshold values. This expectation is corroborated by the top left panel of figure~\ref{fig:alpha_beta_TEB}.}
\label{fig:maps}
\end{figure}

\begin{figure}
  \includegraphics[height=5.cm,width=7cm]{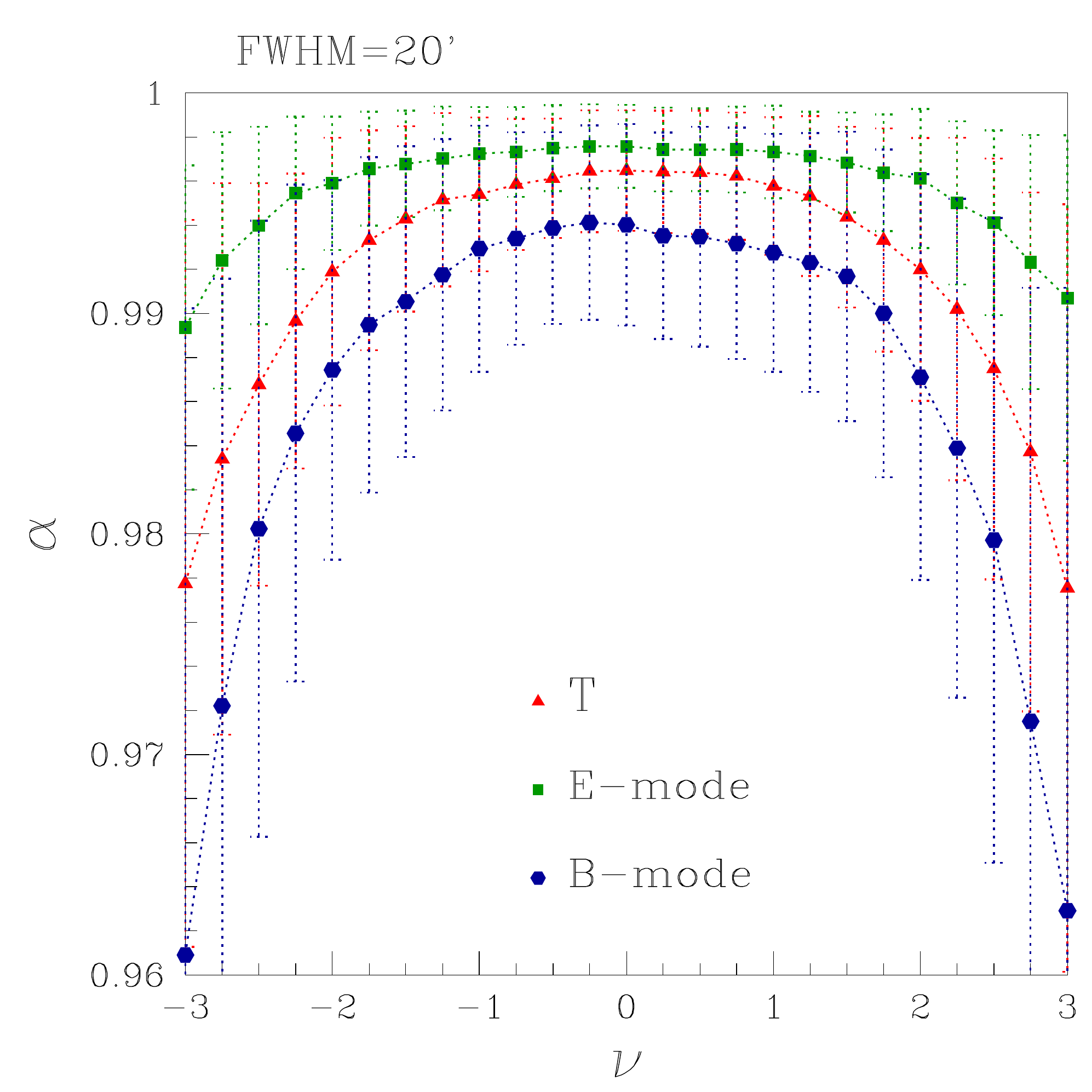}
  \includegraphics[height=5.4cm,width=7cm]{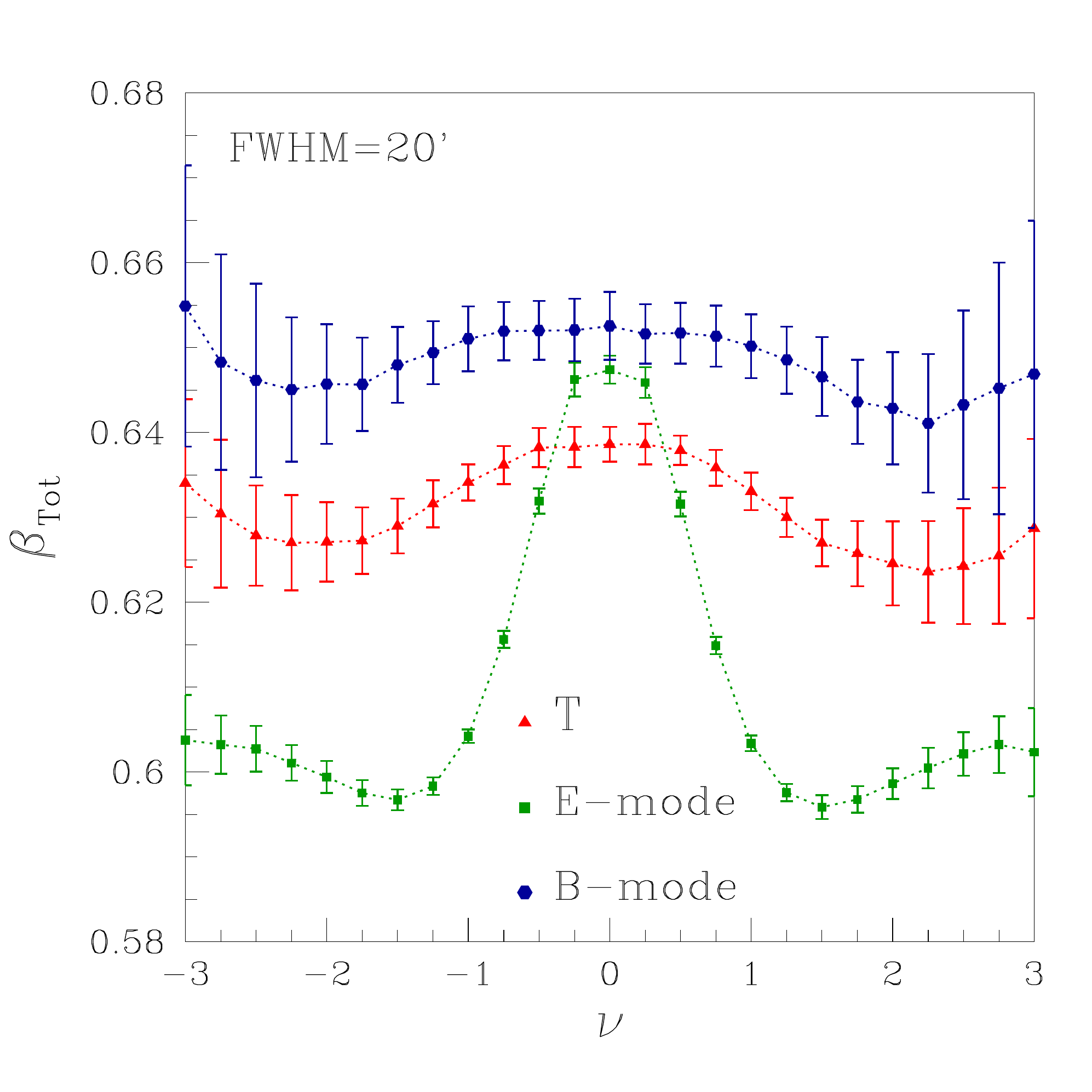}
  \includegraphics[height=5.3cm,width=7.1cm]{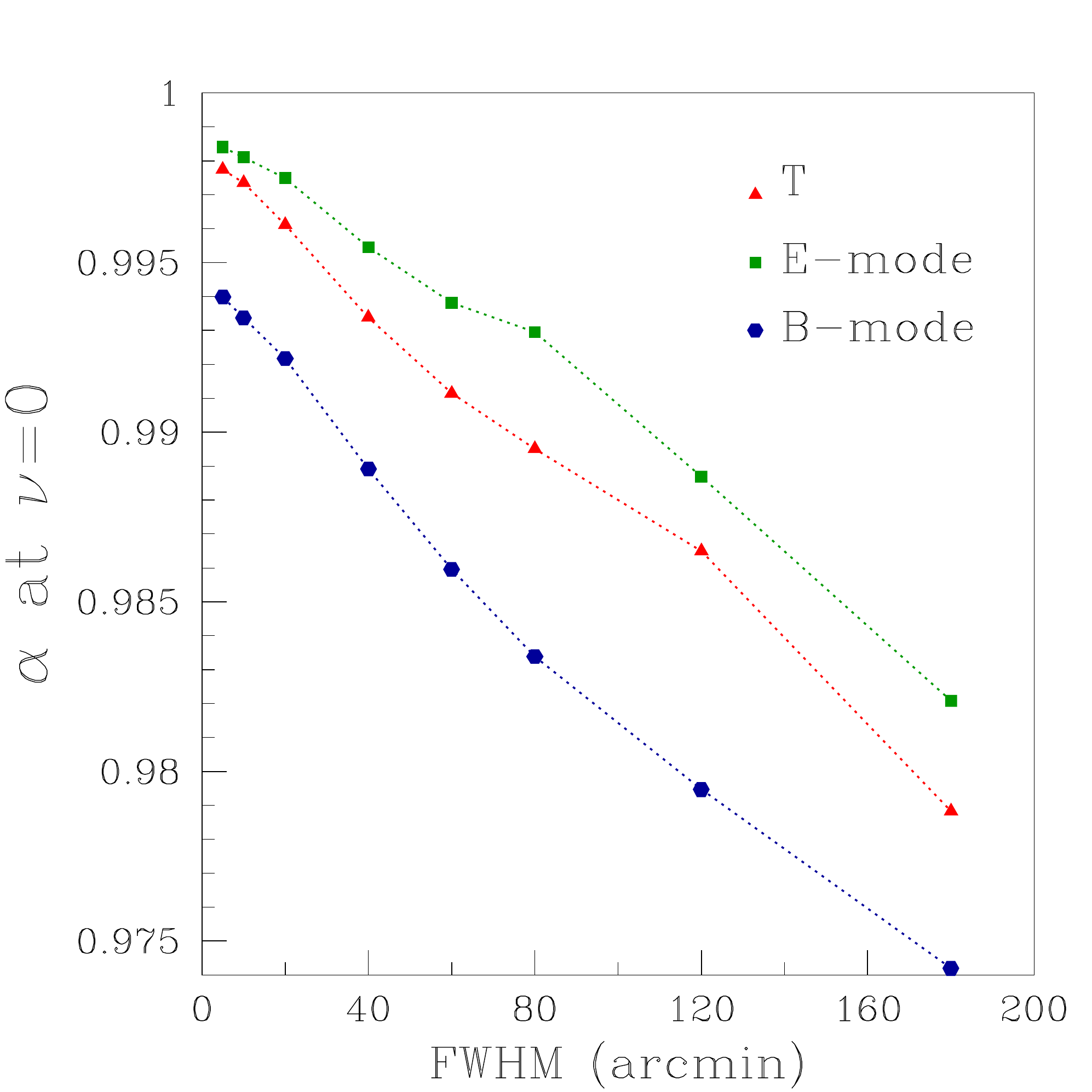}\hskip 1.3cm
  \includegraphics[height=5.3cm,width=7cm]{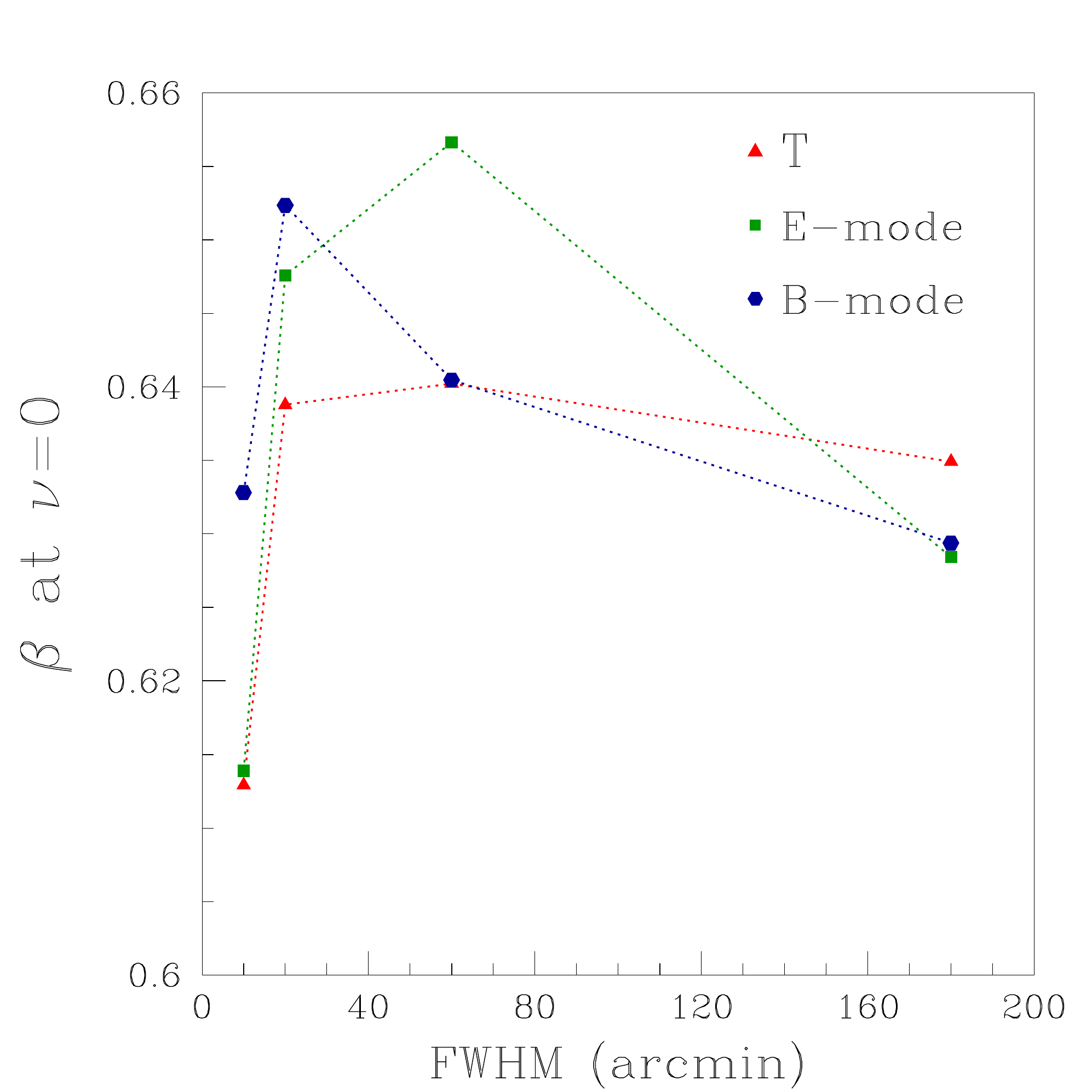}
\caption{{\em Top left}: $\alpha$  versus $\nu$ for $T$, $E$ and $B$ fields for the same smoothing angle FWHM=20'. {\em Bottom left}: Variation of $\alpha$  at $\nu=0$ as a function of the smoothing angle. Both the plots are average over 500 maps.  {\em Top right}:  $\beta$  versus $\nu$ for $T$, $E$ and $B$ fields same smoothing scale as for $\alpha$. {\em Bottom right}: Variation of $\beta$  at $\nu=0$ as a function of the smoothing angle. The plots for $\beta$ are average over 200 maps.}
\label{fig:alpha_beta_TEB}
\end{figure}

In order to first have a visual and intuitive understanding of the number of structures and associated contour length for $T,E,B$ fields, for the same values of $\Lambda$CDM parameters, and same smoothing angle FWHM=60', we show one map of each field in figure~\ref{fig:maps}. We can see that $E$ mode fluctuates the most on smaller scales, thus exhibiting a higher number of structures (hotspots and coldspots) than $B$-mode which fluctuates the least. 
Since the value of $\alpha$ is positively correlated with the total perimeter length  we can anticipate that $\alpha$ will be highest for $E$ mode and lowest for $B$ mode, at all threshold values. 
The plots of $\alpha$ versus threshold, $\nu$, for simulated Gaussian isotropic CMB fields are shown in the top left panel of figure~\ref{fig:alpha_beta_TEB}, for the same values of $\Lambda$CDM parameters, and same smoothing angle FWHM=20'.  The plots are average over 500 realizations.   We can see that $\alpha$ for all three fields are symmetric about $\nu=0$. $E$-mode has the largest values while $B$-mode is the smallest, in agreement with our expectation from the structure of the fields. Here, the value of the tensor-to-scalar ratio is $r=0.1$. Note that change of $r$ will not change the result for $\alpha$ due to the fact that it scales only the field values by the same factor at all pixels, and we use normalized fields $u$ for our calculations. Further, from the bottom left panel showing $\alpha$ at $\nu=0$ as a function of the smoothing scale, we see that $\alpha$ vary monotonically (almost linearly) with smoothing scale, while the slopes are slightly different for the three fields. 
Approximate linear fitting of the dependence of $\alpha$ at $\nu=0$ on the smoothing scale gives the slopes to be $-1.05\times 10^{-4}$ for $T$,  $-9.18\times 10^{-5}$ $E$ mode, and  $-1.16\times 10^{-4}$ for $B$ mode. Thus, $B$ mode has the steepest slope while $E$ mode has the shallowest. 

The plots of $\beta$ versus $\nu$ for the same simulated maps and smoothing scale as for $\alpha$ are shown in the top right panel of figure~\ref{fig:alpha_beta_TEB}.  Note that the $\beta$ values shown here are the average over all structures identified at each $\nu$. These plots are average over 200 realizations\footnote{We use only 200 maps and not 500 as done for $\alpha$ because calculation of $\beta$ is much more computationally expensive.}. $\beta$ for all three fields are symmetric about $\nu=0$. The shapes are quite different. $E$-mode exhibits relatively large variation of $\beta$ across the threshold range, while the structures of $B$-mode are found to be the most isotropic across the entire threshold range. Again, by the same argument as for $\alpha$, $\beta$ for $B$ mode will not change with $r$. The bottom right panel shows $\beta$ at $\nu=0$ as a function of the smoothing scale. We find that $\beta$ does not show monotonic dependence on smoothing scale as $\alpha$. We will further explore the dependence of the shape of $\beta$ versus $\nu$ on the smoothing scale in section~\ref{sec:sec4}. 

\subsection{Sensitivity of $\alpha$ and $\beta$ to variation of cosmological parameters}
\label{sec:alpha_lcdm}

To be able to cleanly isolate the effect of weak gravitational lensing on the morphology of the CMB, as encoded in $\alpha$ and $\beta$, it is important to first understand the effects of various physical parameters which can be potentially confused with lensing. The effects of instrumental noise and residual foreground contamination were studied in~\cite{Joby:2018}. 
For Gaussian fields, the physical properties of the system, or in the context of cosmology the cosmological parameters, enter the scalar/tensorial Minkowski Functionals in the amplitudes and are independent of the threshold dependence. The statistics of interest here, $\alpha$  and $\beta$, are ratios of eigenvalues which are non-linear functions of the matrix elements of $\mathcal{W}_1$. For $\alpha$ it is reasonable to expect that the dependence on cosmology will `mostly' cancel out, leaving behind a weak dependence arising from the non-linear terms of the eigenvalues. However, for $\beta$ it is not clear what to expect since it measures the anisotropy of individual structures.

We have computed $\alpha$ and $\beta$, for $T$, $E$ and $B$, for different sets of cosmological parameters. We find that $\alpha$ is relatively insensitive to variation of cosmological parameters, as anticipated above. $\beta$, on the other hand, is found to have characteristic dependence on $\nu$ for varying cosmological parameters. The details are presented in appendix~\ref{sec:alpha_vary_lcdm}.


\subsection{Sensitivity of $\alpha$ and $\beta$ to Hemispherical anisotropy}
\label{sec:aniso}

We next consider the case of anisotropic CMB temperature field modeled as
\begin{equation}
  \tilde{T}(\hat n) = T^0(\hat n)\left(1+A\, \hat k\cdot\hat n\right),
  \label{eqn:dT_aniso}
 \end{equation}
where $\hat n$ is the pixel direction, $\hat k$ is the direction of anisotropy, and $A$ is the amplitude of the anisotropy. $T^0$ is the isotropic part of the field. This is the so-called dipolar modulation model~\cite{Eriksen:2003db} of hemispherical anisotropy in the CMB. 

We have computed $\alpha$ and $\beta$ or different values of $A$ and different smoothing scales. We find that $\alpha$ is insensitive to it at small smoothing scales, and the input hemispherical anisotropy becomes noticeable in the threshold dependence of $\alpha$ as we increase the smoothing scale. $\beta$, on the other hand, is found to have characteristic dependence on $\nu$. The details are presented in appendix~\ref{sec:alpha_ha}.


\section{Effect of weak lensing on the morphology of CMB fields}
\label{sec:sec4}

Having so far understood the dependence of $\alpha$ and $\beta$ on the smoothing scale and cosmological parameters, and the effect of hemispherical anisotropy, we are now in a position to probe the effect of weak gravitational lensing. We first give a brief introduction to the physics of CMB lensing and the method of simulations of lensed CMB maps, followed by a description of the results.

\subsection{Physics of lensing of CMB/Weak Lensing of CMB}
\label{sec:lensingphysics}
The CMB radiation can be completely characterized by three observables, namely,  temperature $T(\hat n)$ and the Stoke's parameters $Q(\hat n), U(\hat n)$. They are random fields on the sphere. The temperature field is a scalar quantity and can be expanded using spin-0 spherical harmonics as,
\begin{equation}
T(\hat n)=\displaystyle\sum_{l=0}^{l_{max}}\displaystyle\sum_{m=-l}^{l} a_{\ell m}\; Y_{\ell m}(\hat n).
\end{equation}
Since $Q$ and $U$ do not transform as a scalars under rotation of coordinates perpendicular to each line of sight, it is convenient to transform them to the so called $E$-mode and $B$-mode fields which are scalars under rotation. To make this transformation, $Q$ and $U$ are combined into two complex variables $P_{\pm}(\hat n)=Q(\hat n)\pm  iU(\hat n)$. $P_{\pm}$ are spin-2 fields and can be expanded using spin-2 weighted spherical harmonics as,
\begin{equation}
P_{\pm}(\hat n)=\displaystyle\sum_{l=0}^{l_{max}}\displaystyle\sum_{m=-l}^{l} a_{\pm2,lm} \;\; _{\pm2}Y_{\ell m}(\hat n).
\end{equation}
Using spin-2 weighted coefficients we can define multipole coefficients $E_{\ell m}$ and $B_{\ell m}$, and use them to construct $E$- and $B$-mode sky maps as,
\begin{eqnarray}
Multipoles&:& E_{\ell m}=-\frac{1}{2}[a_{2,lm}+a_{-2,lm}] \;,\;  B_{\ell m}=-\frac{1}{2i}[a_{2,lm}-a_{-2,lm}]\\
Maps&:& E(\hat n)=\displaystyle\sum_{\ell m} E_{\ell m}\;Y_{\ell m}, \quad B(\hat n)=\displaystyle\sum_{\ell m} B_{\ell m} \; Y_{\ell m}(\hat n).
\end{eqnarray}
What we observe are the lensed CMB photons whose paths have been altered or deflected multiple times by the gravitational field of large scale structure.

Let the deflection caused by lensing for each sky direction $\hat n$ be captured by deflection angle  $\vec{d}(\hat n)$. Then the observed CMB fields in some direction $\hat{n'}$ can be expressed in terms of the field values in the original direction $\hat{n}$ as, 
\begin{eqnarray}
T^{\rm L}(\hat{n'}) &=& T^{\rm UL}(\hat{n}+\vec{d}\,),\\
(Q+iU)^{\rm L}(\hat{n'})  &=& {} (Q+iU)^{\rm UL}(\hat{n}+\vec{d}).
\end{eqnarray}
For small perturbations (perturbations in the linear regime), assuming the Born approximation, the deflection field is given by the gradient of the lensing potential, as $\nabla_{\hat{n}}\Phi$. The lensing potential $\Phi(\theta,\phi)$ is related to the space and time dependent gravitational potential field $\psi$ as,
\begin{equation}
\Phi(\hat{n})=-2\int_0^{\chi^*} \d\chi \left(\frac{\chi^*-\chi}{\chi\chi^*}\right) \psi \left(\chi\hat{n}; \eta_0-\chi\right),
\end{equation}
where $\chi$ is the comoving distance, $\chi^*$ is the comoving distance to the CMB last scattering surface, $\psi$ is the 3-dimentional gravitational potential at conformal distance $\chi$ along the direction $\hat{n}$, and $\eta_0-\chi$ is the conformal time.

The remapping of CMB field values in the sky alters the power spectrum of both temperature and $E$ mode anisotropies by smoothening the acoustic peaks and adding power towards the damping tail~\cite{Seljak:1995ve}. For polarization the most important quantitative effect is the generation of $B$-mode at small angular scales. 
The observed $B$-mode polarization signal has two components. The first is the primordial component sourced by gravitational waves generated during inflation. The amplitude of this component depends on the value of the tensor-to-scalar ratio, $r$. For values of $r$ around the current upper limit of 0.07~\cite{Ade:2018gkx} the primordial component dominates at large angular scales. The  second component, which is the one generated by weak lensing, dominates at small scales

CMB lensing is an integrated effect and hence, sensitive to all the matter along the line of sight, thus acting as a complementary tracer of the large-scale structure in the universe. The lensing potential defined above can be reconstructed from the observed lensed CMB. The reconstructed lensing map can be further used for delensing the observed CMB to extract the primordial unlensed CMB fields.
It is crucial to understand the properties of lensed CMB fields especially the $B$-mode so that in future it can be easily separated from the unlensed primordial $B$-mode. Since, the lensed CMB observables are non-Gaussian, in addition to the power spectrum we need other statistical measures to  study their statistical properties. In the following subsections, we study the changes in the morphology of the CMB fields induced by lensing by using the contour Minkowski Tensor which can capture the information about shapes and the relative alignment of the structures. 

\subsection{Simulation of lensed CMB maps}
\label{sec:simulation}
Weak lensing of CMB can be modelled very well using linear physics and one can simulate full sky lensed CMB maps in weak lensing approximation using \texttt{LENSPIX}~\cite{Lenspix}. Under the weak lensing approximation, this package implements a pixel remapping approach to mimic the effects of lensing on the CMB photons. The temperature or polarization field is shifted by the deflection vector as a function of position on the surface of sphere.
Lensed temperature and polarization fields in a particular direction having coordinates $(\theta,\phi)$ are given by unlensed fields in another direction $(\theta',\phi')$ at the last scattering surface as,
\begin{eqnarray}
\tilde{T}(\theta,\varphi)&=&T(\theta',\varphi'),\\ 
\tilde{P}(\theta,\varphi) &	= & \exp\left[-2{\rm i}(\gamma-\alpha)\right]\; P(\theta',\varphi').
\end{eqnarray}
The angular coordinates corresponding to the original direction of the photon path $(\theta',\varphi')$, are determined by the deflection field $\vec{d}(\theta,\varphi)$,
\begin{eqnarray}
\displaystyle \cos\theta'	&=& 	\displaystyle \cos d~\cos\theta-\sin d~\sin\theta~\cos\alpha, \\
\displaystyle \sin(\varphi'-\varphi)	&=& 	\displaystyle \frac{\sin\alpha~\sin(d)}{\sin\theta'},
\end{eqnarray}
where the deflection vector field is given by
\begin{equation}
\vec{d}\equiv \vec{\nabla} \Phi = d_{\theta}\, e_{\theta} + d_{\phi}\, e_{\phi} = d \, \cos \alpha \, e_{\theta} + d \, \sin \alpha \, e_{\phi}.
\end{equation}

We have generated maps of lensed and unlensed $T,E,B$ using \texttt{Lenspix}. The theoretical input power spectra for lensing potential and unlensed $T$, $E$ and $B$ fields was obtained using the \texttt{CAMB}~\cite{Lewis:2000ah,cambsite} code. 
The input $\Lambda$CDM parameters are: $\Omega_bh^2=0.0226$, $\Omega_{cdm}h^2= 0.1120$, $H_0=70\, {\rm km\,s^{-1}\, Mpc^{-1}}$, $\tau=0.09$, where the symbols have their usual meanings. Unless otherwise mentioned, all $B$ mode maps have input value of tensor-to-scalar ratio $r=0.1$.
For calculations done with smoothing scales 5' and higher we use maps for which the \texttt{HEALPIX} resolution parameter is $N_{\rm side} = 1024$ for both lensed and unlensed case. For large smoothing angles we downgrade the maps to appropriate $N_{\rm side}$ values before smoothing them.  
For smoothing scales lower than 5' we use $N_{\rm side} = 2048$.
Then we numerically compute the CMT for each smoothed map using the methods described in section~3. We rescale the field by the corresponding rms value so the typical threshold value is of order one. We choose the threshold
range -$ 4.0 < \nu < 4.0$ with $33$ equally spaced bins for our calculations. 

All error bars shown in this paper are the sample variance obtained from the number of maps of each field that is used. For $\alpha$ error bars are calculated using 500 simulations each of unlensed and lensed fields. The calculation of $\beta$ is much more computationally intensive. Hence we use 200 simulations. For our study, we consider an ideal case ignoring contamination from instrumental noise, foreground emissions and other effects.

\subsection{Morphological changes induced by lensing}

Since lensing results in remapping of the CMB field values at different pixels we  expect that it will lead to distortions of hotspots and coldspots. 
However, statistical isotropy is expected to be preserved provided the distribution of the lensing potential remains isotropic. We quantify below the effect of lensing on the statistical isotropy parameter $\alpha$, and the distortions induced for individual structures quantified by $\beta$. 
We note that while $T$ is directly remapped by lensing,  $E$ and $B$ are obtained from  the directly remapped polarization components $Q,U$ via non-local relations~\cite{Zaldarriaga:2001st}. Hence, we do not expect similar  distortion effects for all the fields. 

\begin{center}
\begin{figure}
\includegraphics[height=2.in,width=2.0in]{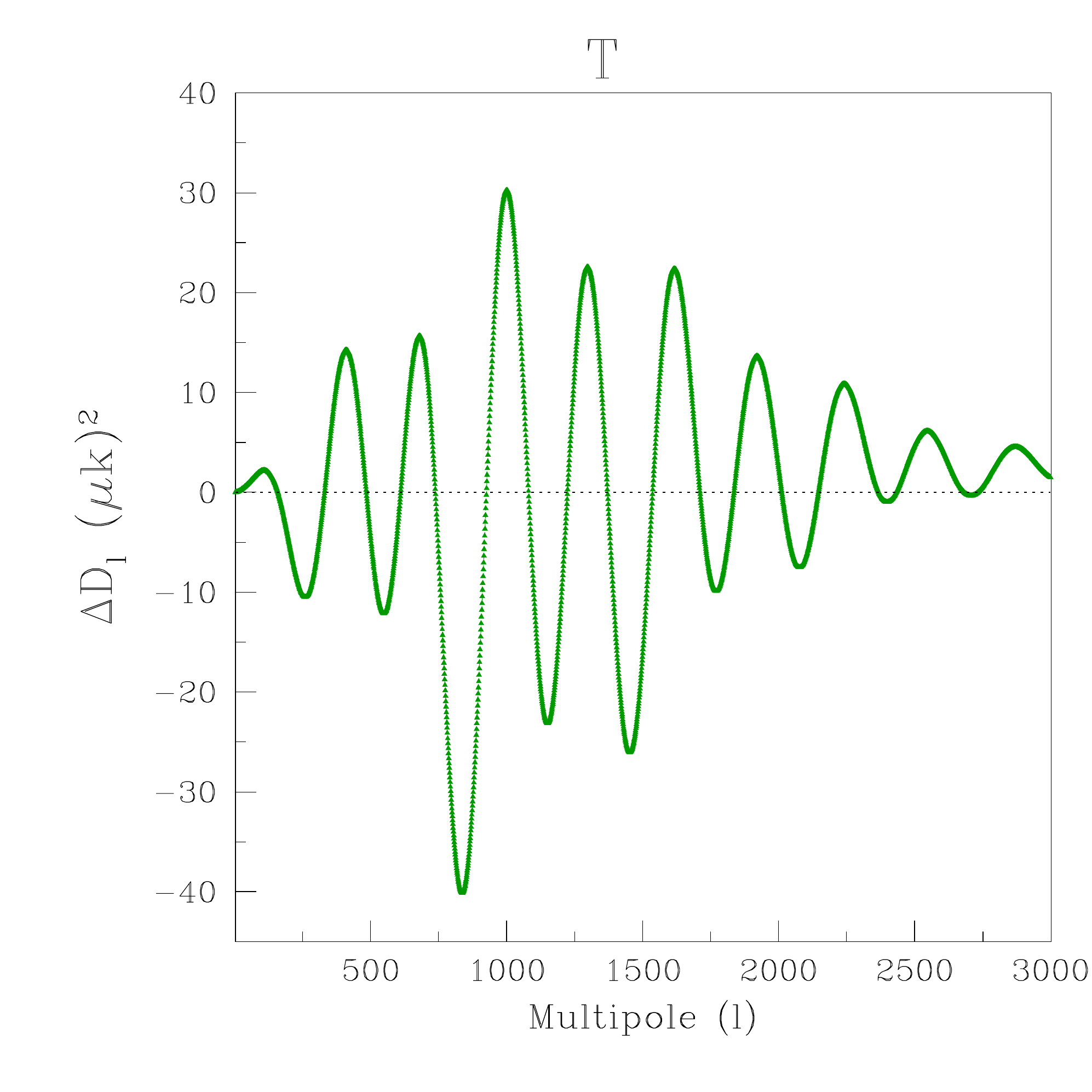}
\includegraphics[height=2.in,width=2.0in]{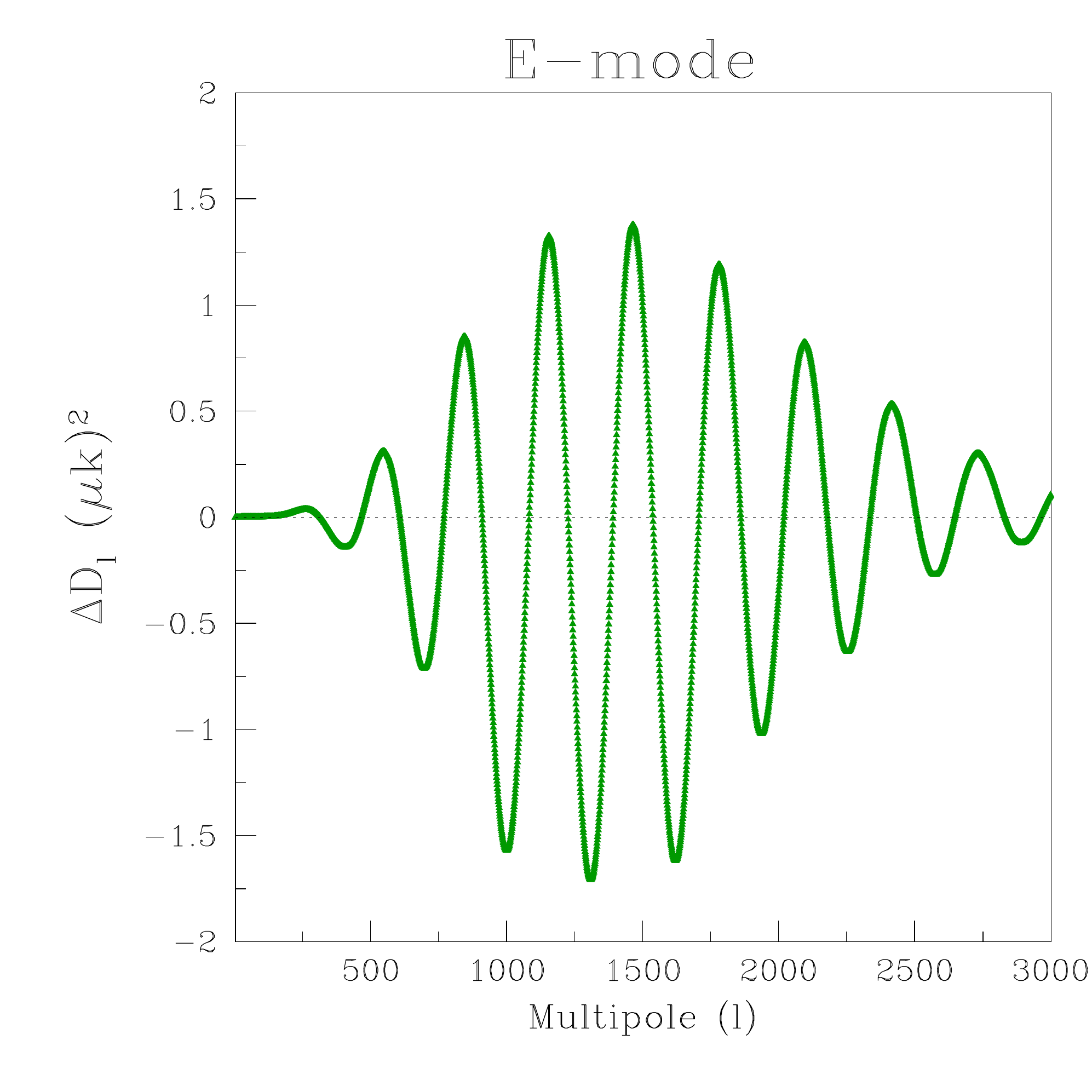}
\includegraphics[height=2.in,width=2.0in]{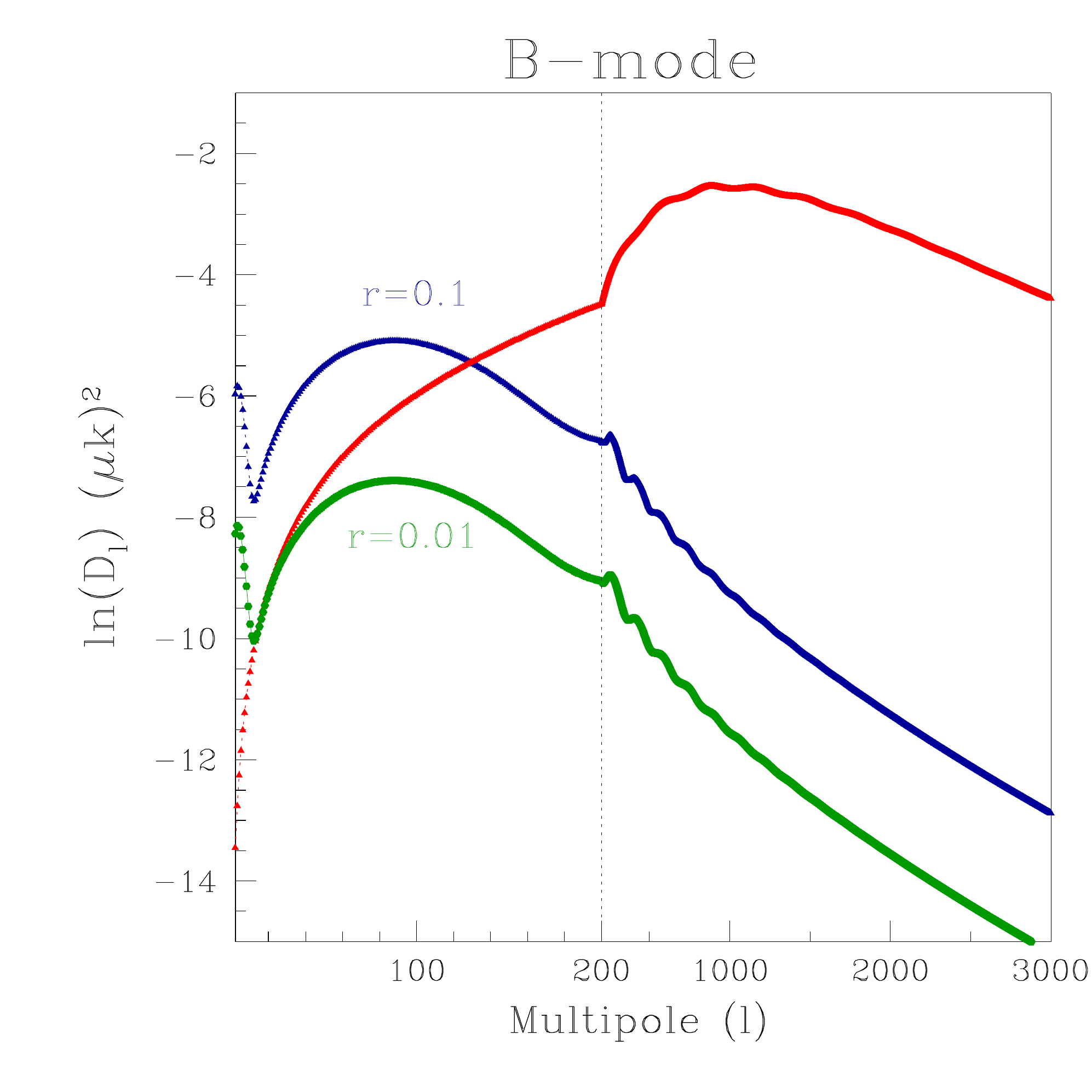}
\caption{{\em Left}: Difference between lensed and unlensed angular power spectra for $T$ used in our simulations.  {\em Middle}: same for $E$.  {\em Right}: $D_{\ell}$ for primordial $B$ with $r=0.1$ (blue) and $r=0.01$ (green), and lensed $B$ (red). We have used two different scales on the $x$-axis, with a dotted line serving as demarcation, in order to highlight the low $\ell$ region.}
\label{fig:cl_TEB}
\end{figure}
\end{center}

The left and middle panels of figure~\ref{fig:cl_TEB} show the difference of the lensed and unlensed angular power spectra, $\Delta D_{\ell}$, where $D_{\ell}\equiv \ell(\ell+1) C_{\ell}/2\pi$, for $T$ and  $E$. The right panel shows $D_{\ell}$ for $B$ mode for unlensed (primordial) case with $r=0.1$ (blue) and $r=0.01$ (green), and  lensed case (red).  These plots serve the purpose of checking our simulations by comparing with known results, such as in~\cite{Lewis:2000ah}. For $B$ mode the amplitude of the unlensed power  decreases linearly with $r$. For $r=0.1$ we can see that the power of the lensed component dominates over the primordial component at scales $\ell \gtrsim 150$ or larger than degree angular scales. For smaller values of $r$ the transition will take place at even larger angular scales. In the following subsections we present our findings on the morphological changes of the CMB fields induced by lensing. We will find it useful to compare with the lensing effects on the power spectra when we interpret our results on the morphological changes.

\subsubsection{Effect of lensing on $\alpha$}

We compute $\alpha$ for unlensed and lensed maps for the three fields - temperature, $E$ and $B$ modes. The calculations are carried out for different smoothing scales in order to find out how the effect of lensing varies with scale. Let us denote the difference of $\alpha$ between the lensed and unlensed maps by  
\begin{equation}
  \Delta \alpha(\nu) =   \alpha^{\rm L}(\nu) - \alpha^{\rm UL}(\nu).
\end{equation}
\begin{center}
\begin{figure*}
\includegraphics[height=2.6in,width=3in]{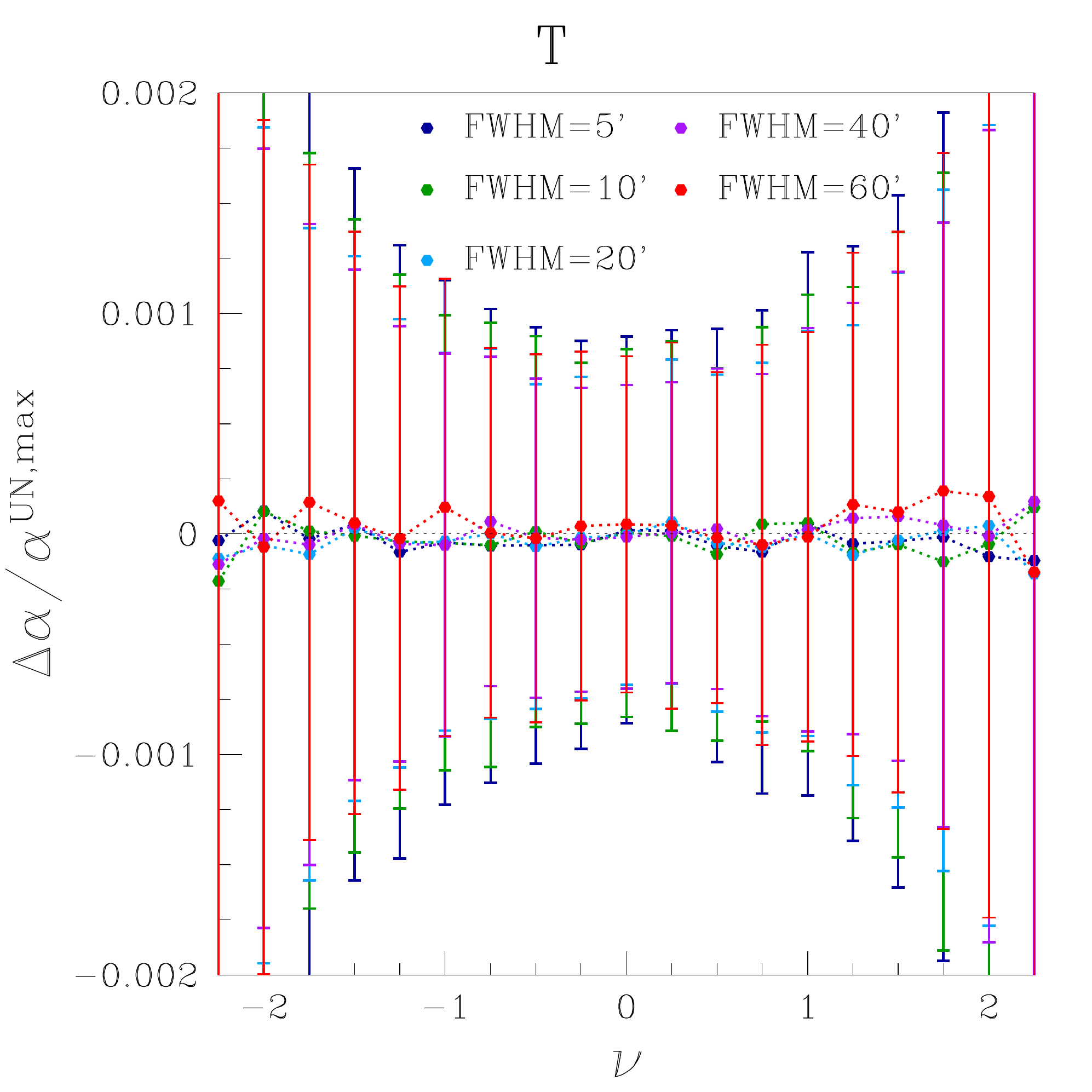}
\includegraphics[height=2.6in,width=3in]{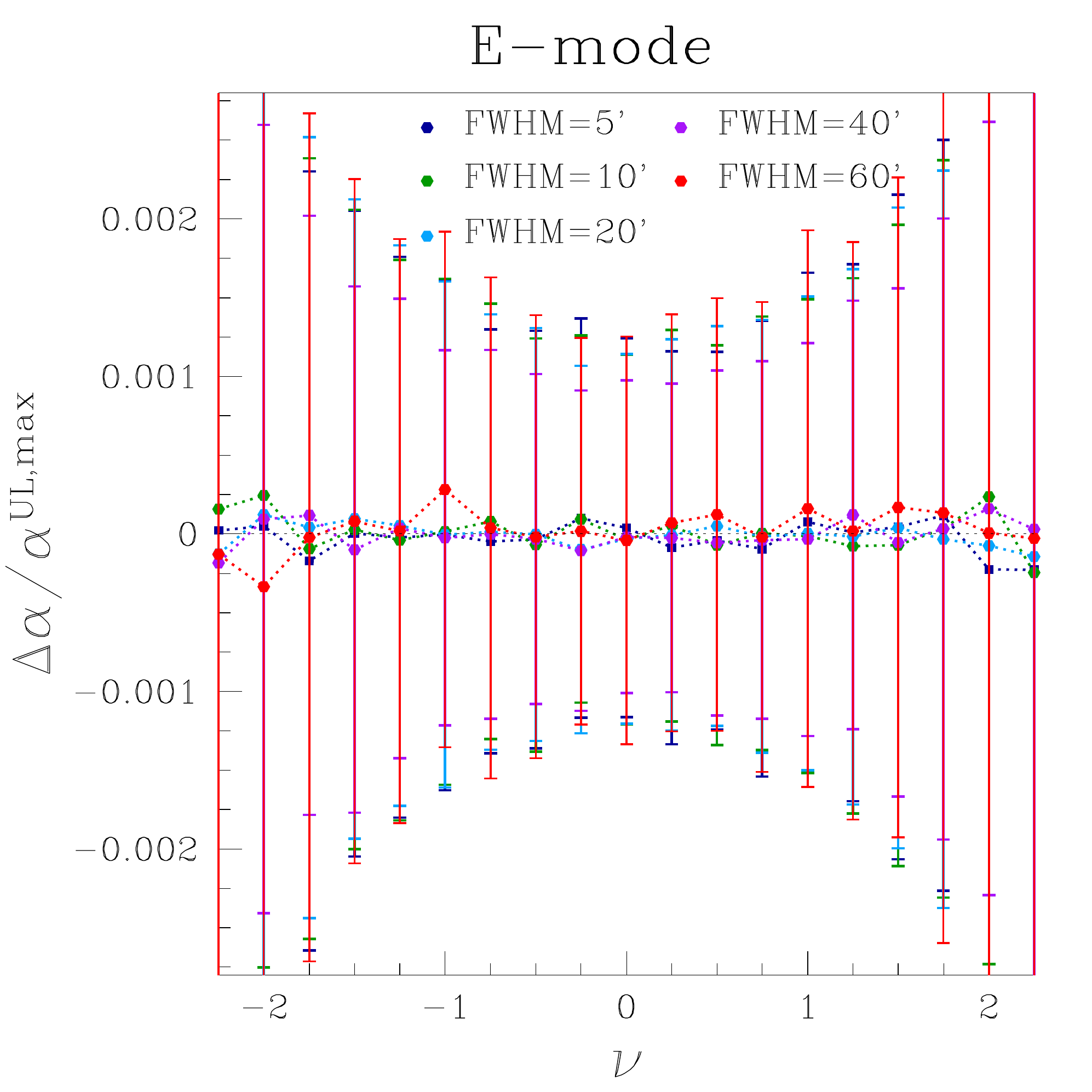}
\includegraphics[height=2.6in,width=3in]{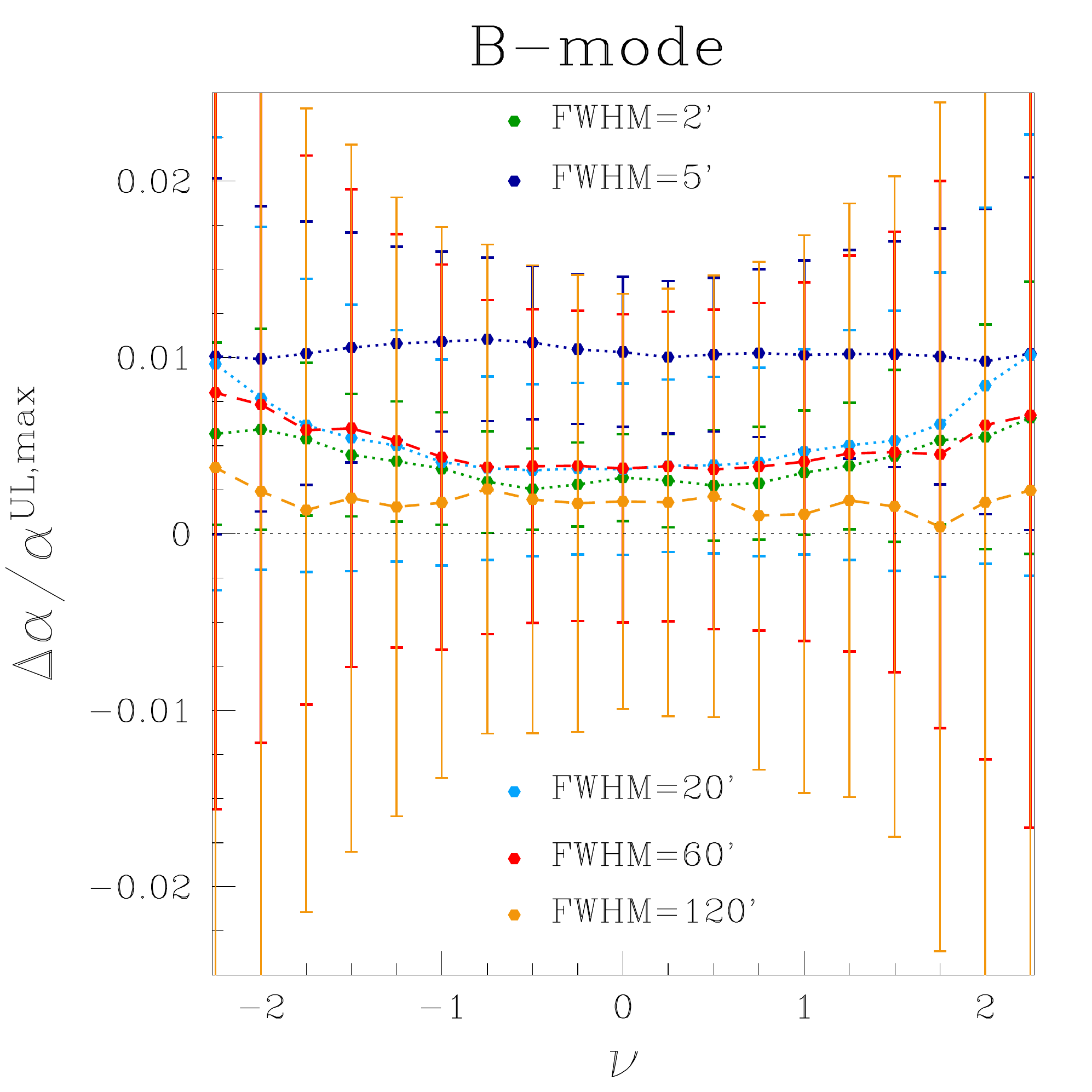}
\includegraphics[height=2.6in,width=3in]{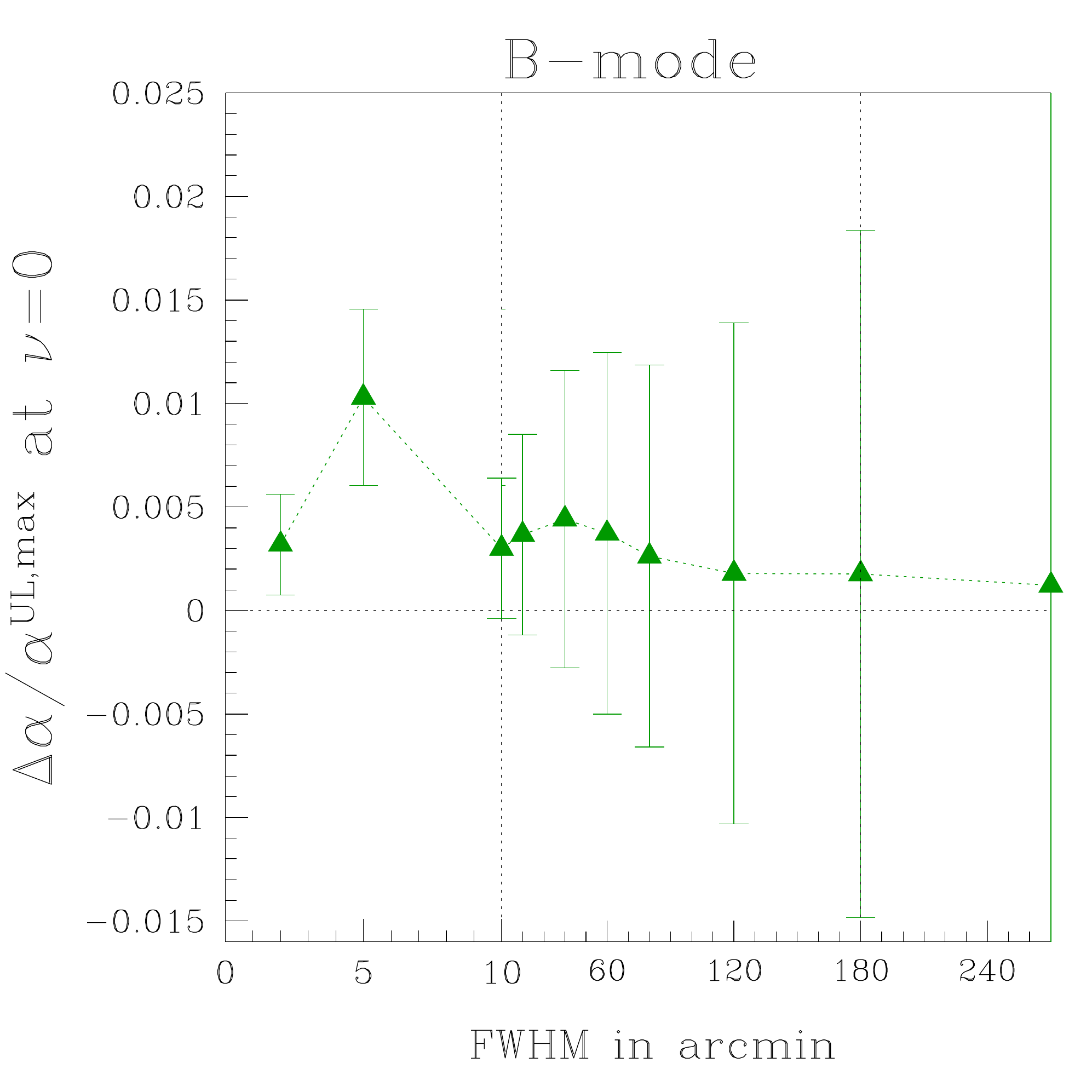}
\caption{{\em Top panels}: Deviations $\Delta\alpha$ between lensed and unlensed $T$ (left), $E$-mode (right). {\em Bottom left panels}: $\Delta\alpha$ for $B$-mode. All plots are average over 500 maps and
  the error bars are the standard deviation obtained from these
  maps. Note that the smoothing angles are different for the different
  fields. 
  {\em Bottom right:}  $\Delta\alpha$ at $\nu=0$ as a function of smoothing angle, for $B$-mode, normalized by the magnitude of unlensed $\alpha$ at $\nu=0$. The range of the $x$-axis that is probed is between FWHM=$2'$ to 270'. This range is subdivided into three different scales so as to highlight the behaviour at different scales. The locations of the scale transitions at FWHM=10' and 180' are marked by vertical dotted lines. Note the peak at FWHM$\sim 5'$.}
\label{fig:alpha_TEB}
\end{figure*}
\end{center}
Let $\alpha^{\rm UL,max}$ denote the magnitude of the unlensed $\alpha$ at $\nu=0$ where $\alpha$ has the maximum value. 
The top panels of figure~\ref{fig:alpha_TEB} show $\Delta\alpha$ normalized by  $\alpha^{\rm UL,max}$ as a function of $\nu$ at various smoothing scales, for temperature and $E$ mode fields. The plots are averaged over 500 realizations. We find $\Delta\alpha$ for $T$ and $E$  fluctuate about zero with error bars which show consistency of  $\Delta\alpha$ being zero at all $\nu$, for all the smoothing scales considered here. Therefore, we do not obtain any systematic variation of $\Delta \alpha$ with smoothing scale.

Next we consider the effect of lensing on $B$ mode field. From the right panel of figure~\ref{fig:cl_TEB} we see that the angular power spectrum of the primordial $B$ mode peaks at around $\ell \sim 80$ which corresponds to roughly $2^{\circ}$ angular scale. Beyond this scale the power dissipates rapidly towards small angular scales. 
The lensing induced power leakage from $E$ to $B$ peaks at around $\ell\sim 1000$. For $r=0.1$ the lensing power dominates at scales smaller than $1^{\circ}$. This is manifest as structures in the $B$ mode field at corresponding angular scales. Consequently, $\alpha$ for lensed $B$ is expected to be larger than that of the unlensed value at all threshold values at scales dominated by the lensed signal. This is what we obtain, as seen in the bottom left panel of figure~\ref{fig:alpha_TEB}, where the primordial $B$ mode maps have input $r$ value $0.1$. We find $\Delta\alpha$ is positive for all the smoothing scales that we have considered here. Further, we find that the effect of lensing becomes more pronounced as $|\nu|$ increases. The error bars also increase as $|\nu|$ increases since the total perimeter of excursion sets drops exponentially with increasing $|\nu|$. The functional form of the variation of $\Delta\alpha$ with the smoothing scale is, however, not clear from this plot.  

To get a clear picture of how $\Delta\alpha$  varies with the smoothing scale we show the normalized $\Delta\alpha$ at $\nu=0$ versus smoothing angle in the bottom right panel of figure~\ref{fig:alpha_TEB}. Note that on the $x$-axis we have used three different linear scales with transitions at FWHM=10' and 180'. In the range between 2' and 10' we find that $\Delta\alpha$ peaks around FWHM=5' which corresponds to a smoothing angle $\theta_s \sim 5'/2\sqrt{2\ln 2} \sim 2.12'$ . This corresponds to the characteristic angular scale set by the typical size of the lensing potential. Notice that the size of the error bars are smaller towards smaller angular scales which indicates that the total perimeter of the structures increases towards smaller angles. In the range between 10' and 180' we find that $\Delta\alpha$ exhibits a mild increase that peaks at roughly FWHM=40'. The statistical significance decreases, as seen from the increase of the error bars resulting decrease of the perimeter. Further $\Delta\alpha$ decreases towards large angle as is expected from the fact that the lensing contribution to $B$ mode becomes subdominant in comparison to the primordial part. However, we do find a small but non-zero value even at 270' smoothing which is possibly due to the fact that lensing correlates signals at higher than degree scales.

\subsubsection{Effect of lensing on $\beta$}

\begin{figure}
\includegraphics[height=1.8in,width=2in]{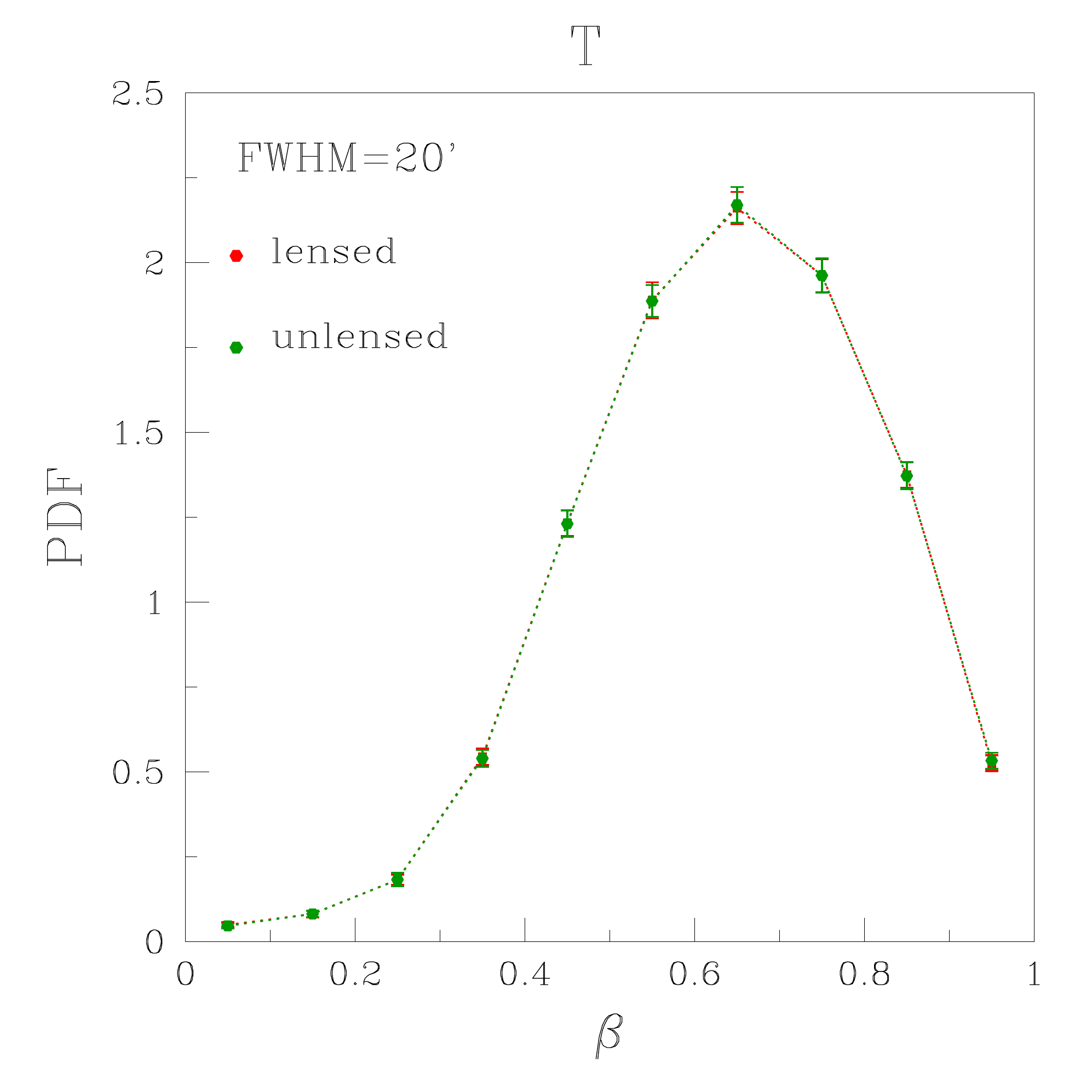}
\includegraphics[height=1.8in,width=2in]{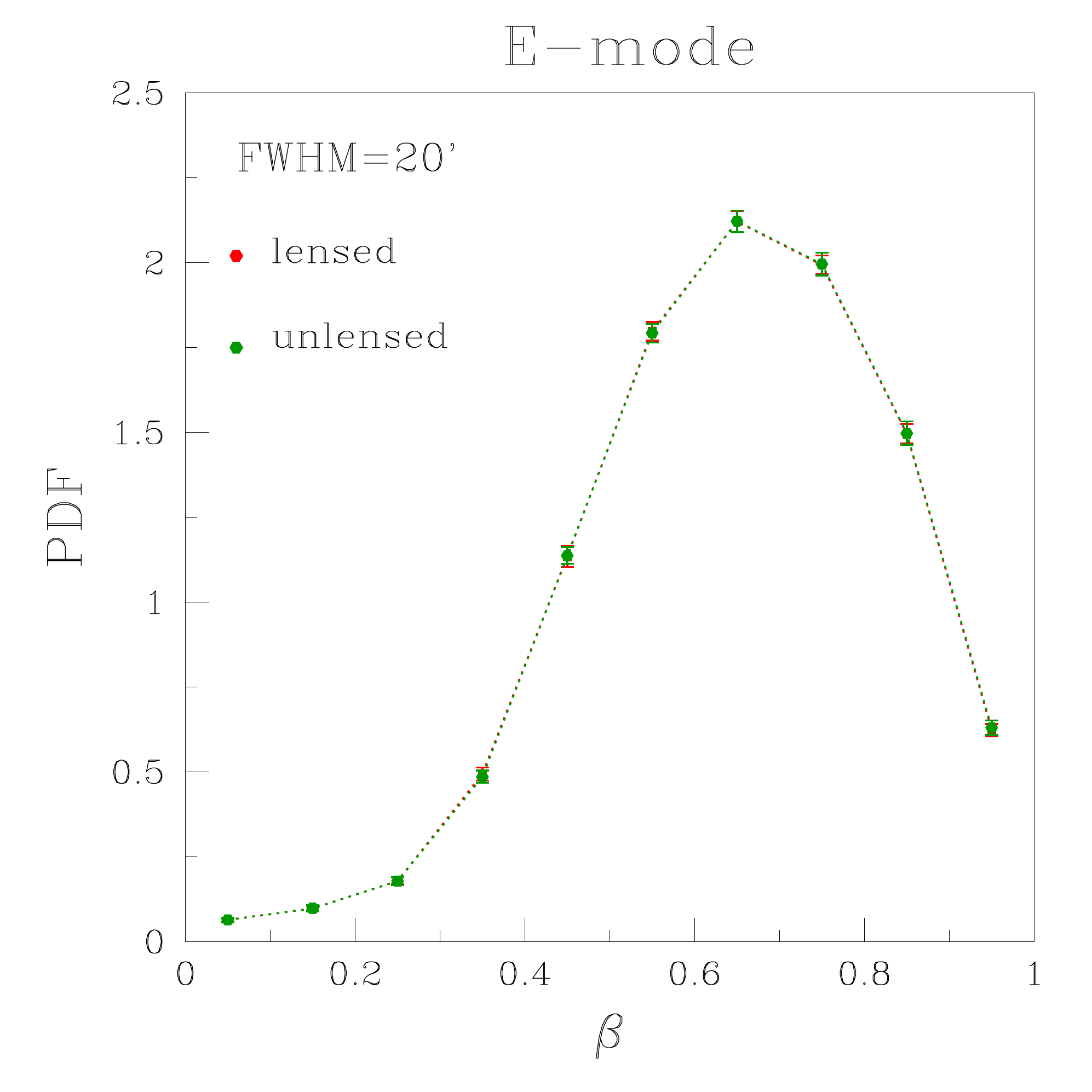}
\includegraphics[height=1.8in,width=2in]{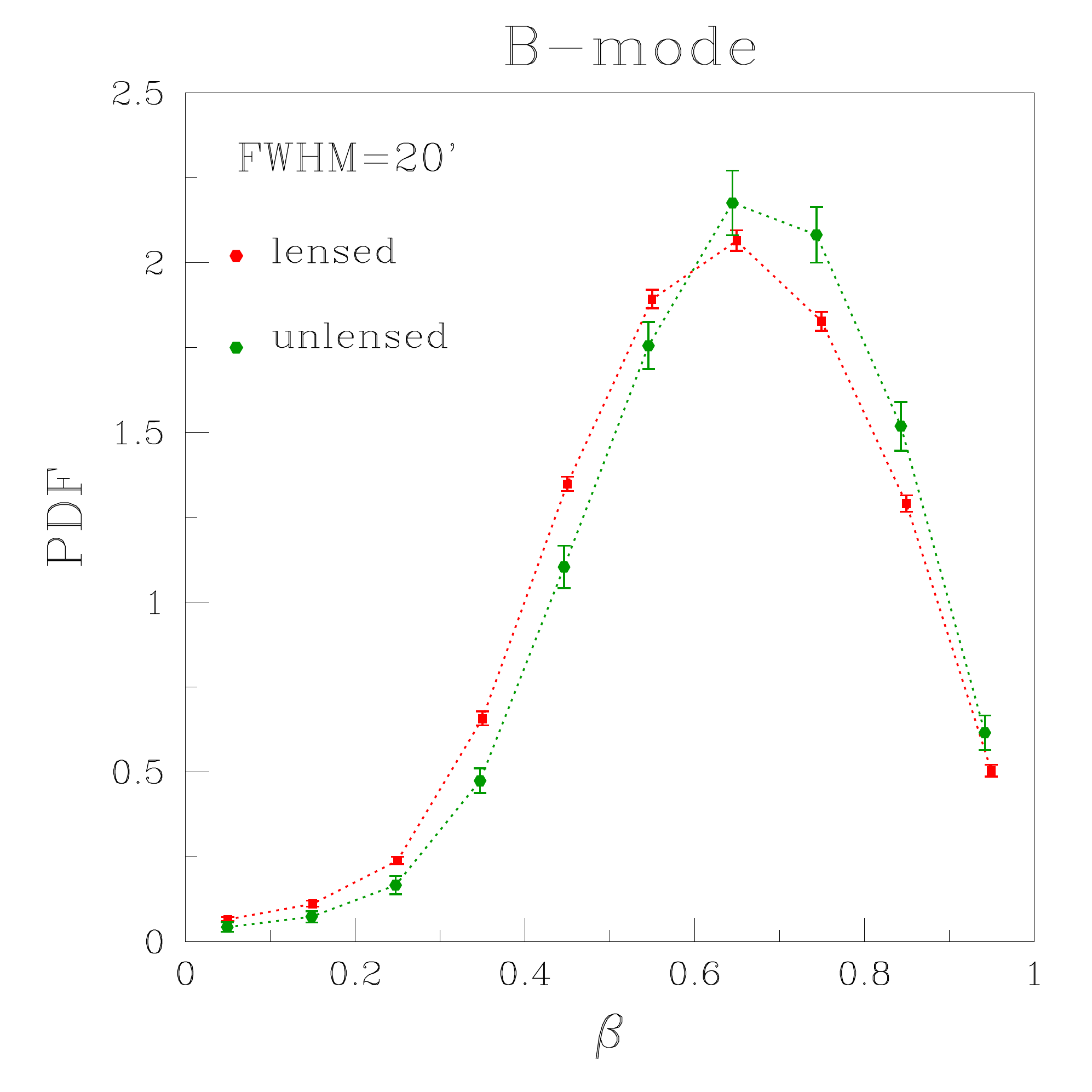}\\
\includegraphics[height=1.8in,width=2in]{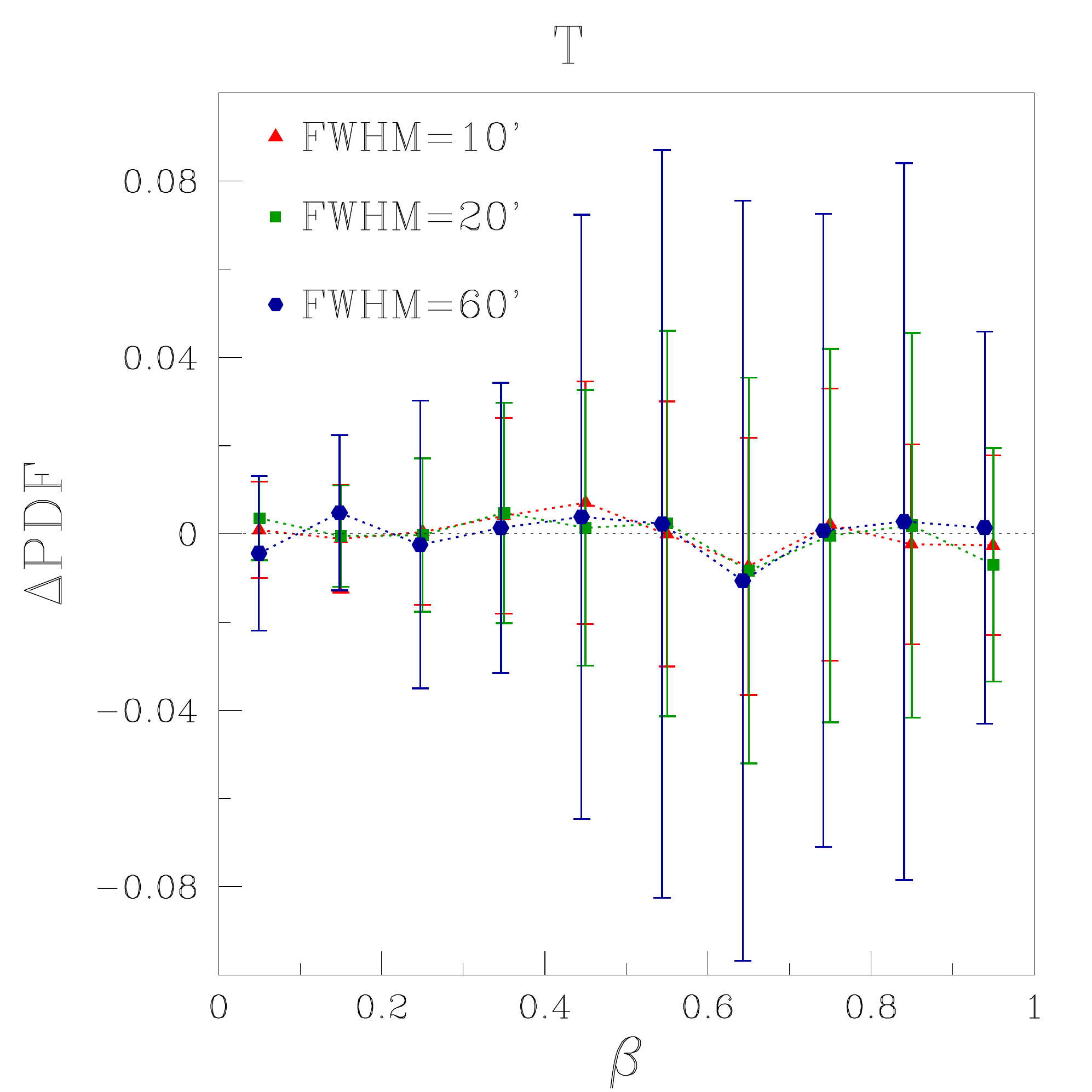}
\includegraphics[height=1.8in,width=2in]{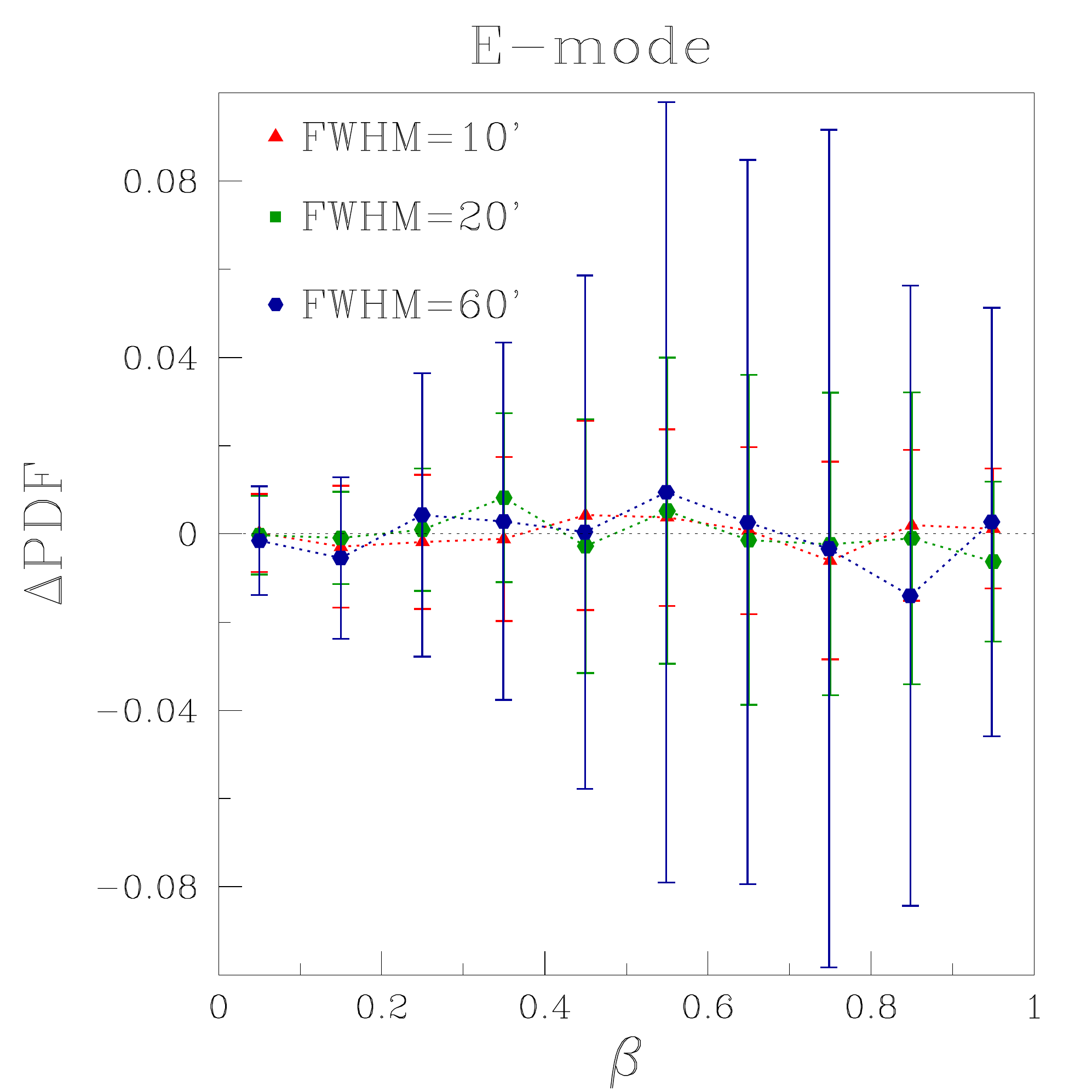}
\includegraphics[height=1.8in,width=2in]{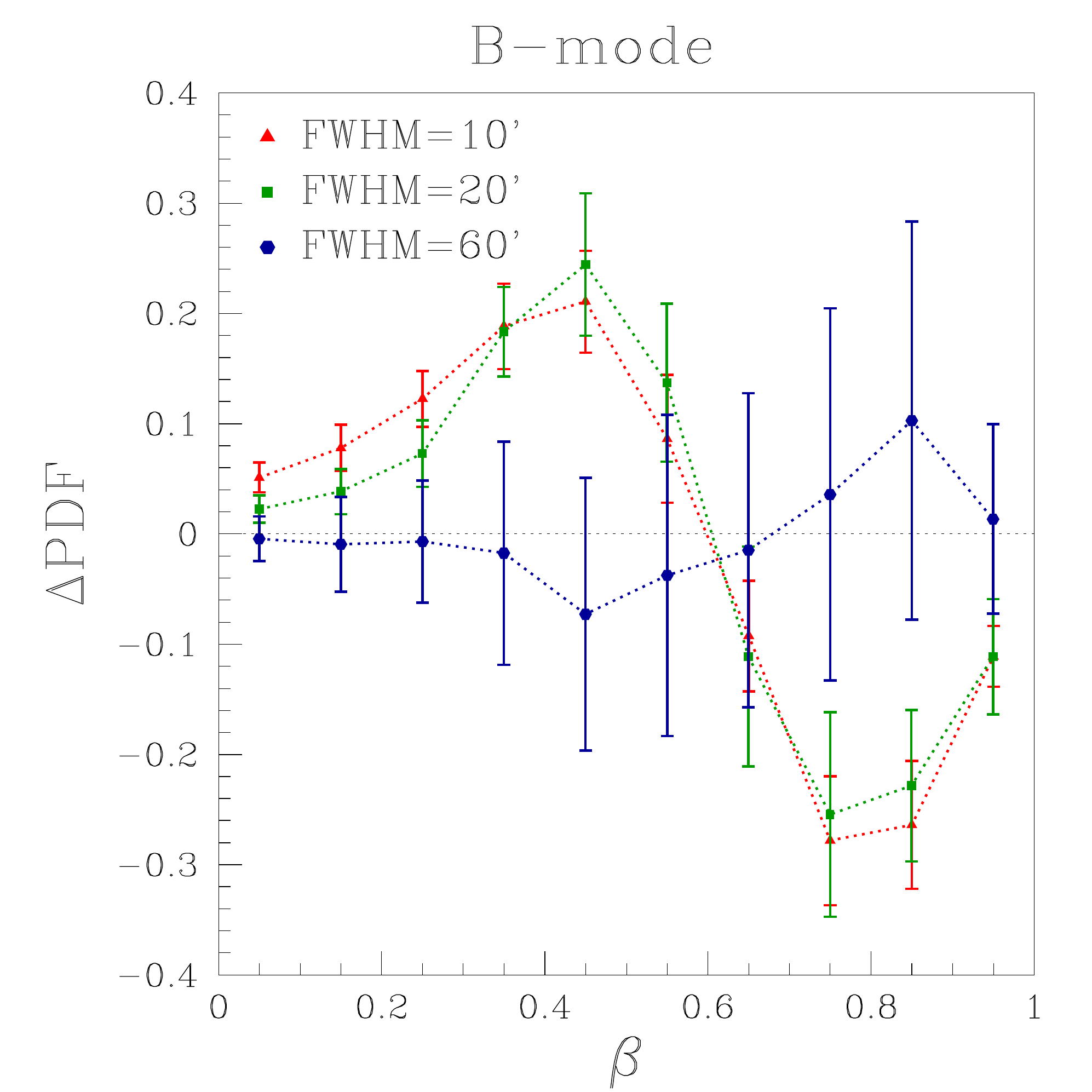}
\caption{{\em Top panels}: PDFs of $\beta$ for $T$ (left), $E$ (middle) and $B$ (right), at $\nu=0$ for FWHM=20' for unlensed (green) and unlensed (red) cases. {\em Bottom panels}: The difference between lensed and unlensed PDFs for the three fields shown in the top panels. All plots are average over 200 maps and error bars are the standard deviations obtained from them.}
\label{fig:pdf_of_beta}
\end{figure}

Next we quantify the distortion induced by lensing on individual structures. This information is captured by the $\beta$ statistic. Again,  the distortion effects are expected to be different for $T$ and $E$ and $B$ fields because $T$ is directly remapped by lensing, while for $E$ and $B$ the effect is indirect via $Q,U$.

In the top panels of figure~\ref{fig:pdf_of_beta} we show the probability distribution function (PDF) of $\beta$ obtained using all structures (hotspots and coldspots) at $\nu=0$. The PDFs for lensed maps is shown in red and for unlensed in green. The left panel shows $T$, middle shows $E$ and right shows $B$, for smoothing $FWHM=20'$. The corresponding bottom panels show the difference between lensed and unlensed PDFs for three smoothing scales given by $FWHM=10',\ 20', \ 60'$. We do not find noticeable effect on the PDFs for $T$ and $E$ at the smoothing scales we have considered. 
$B$ mode, on the other hand, exhibits a relative shift of the peaks of the the PDFs, indicating that lensed structures are more anisotropic.

\begin{center}
\begin{figure}
\includegraphics[height=2.4in,width=2in]{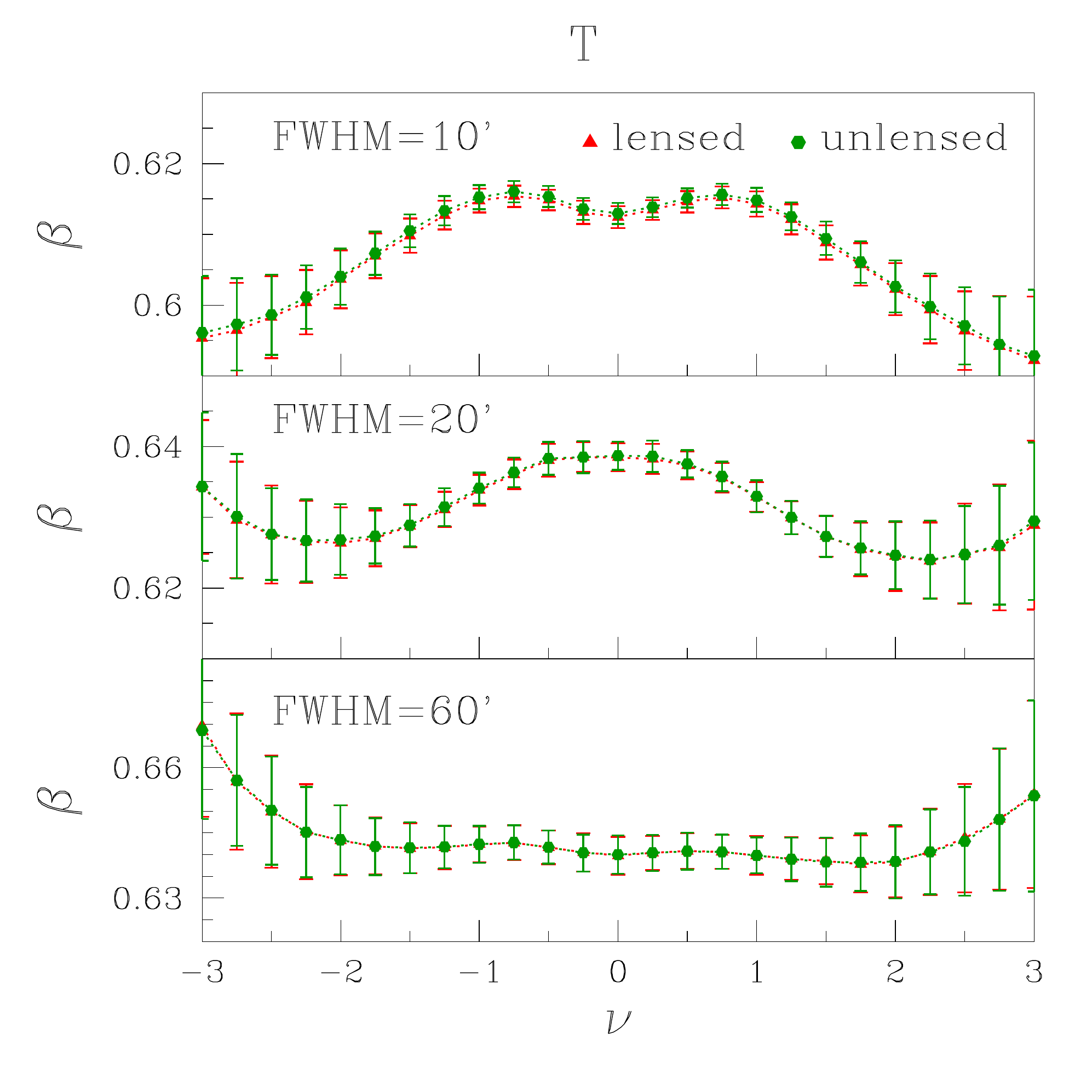}
\includegraphics[height=2.4in,width=2in]{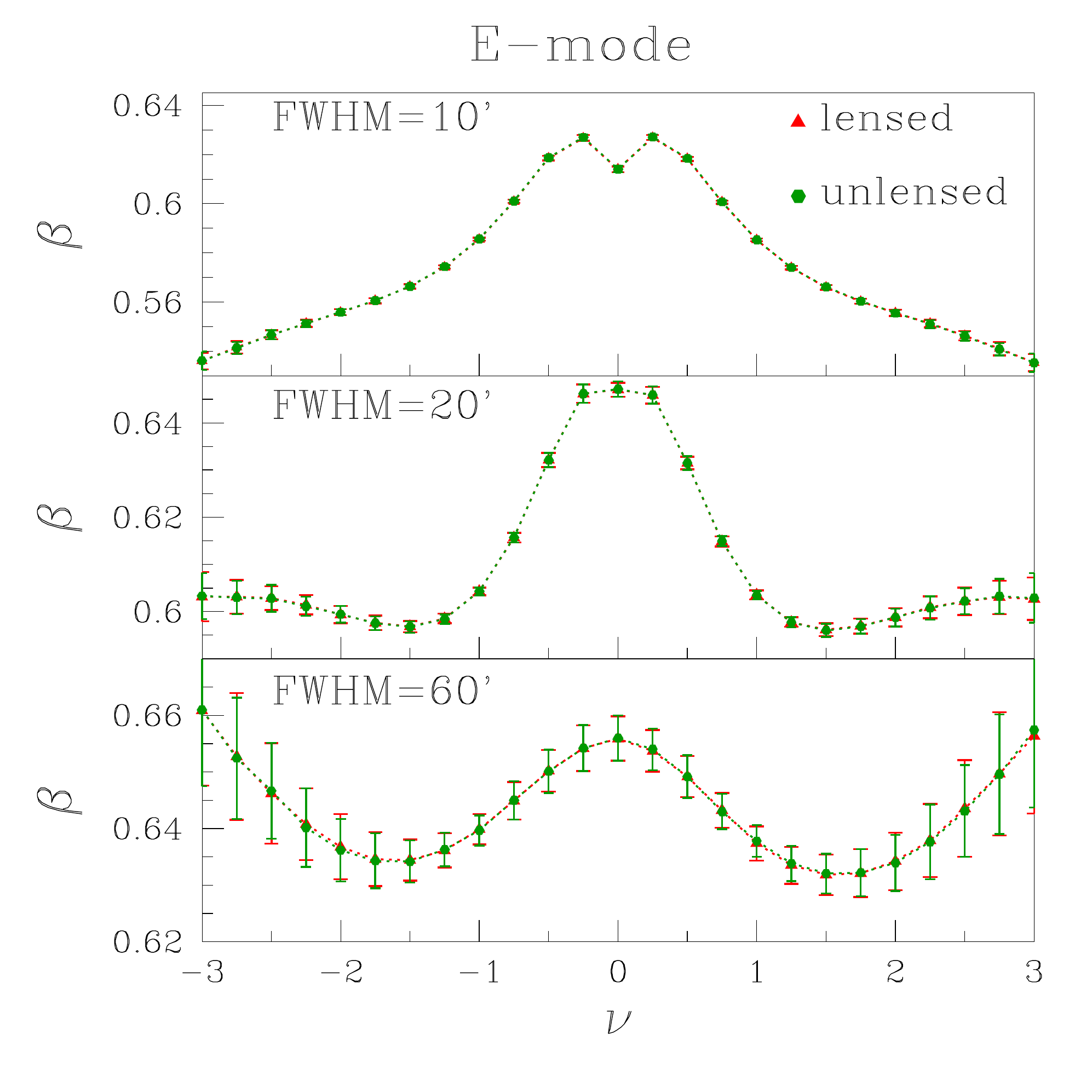}
\includegraphics[height=2.4in,width=2in]{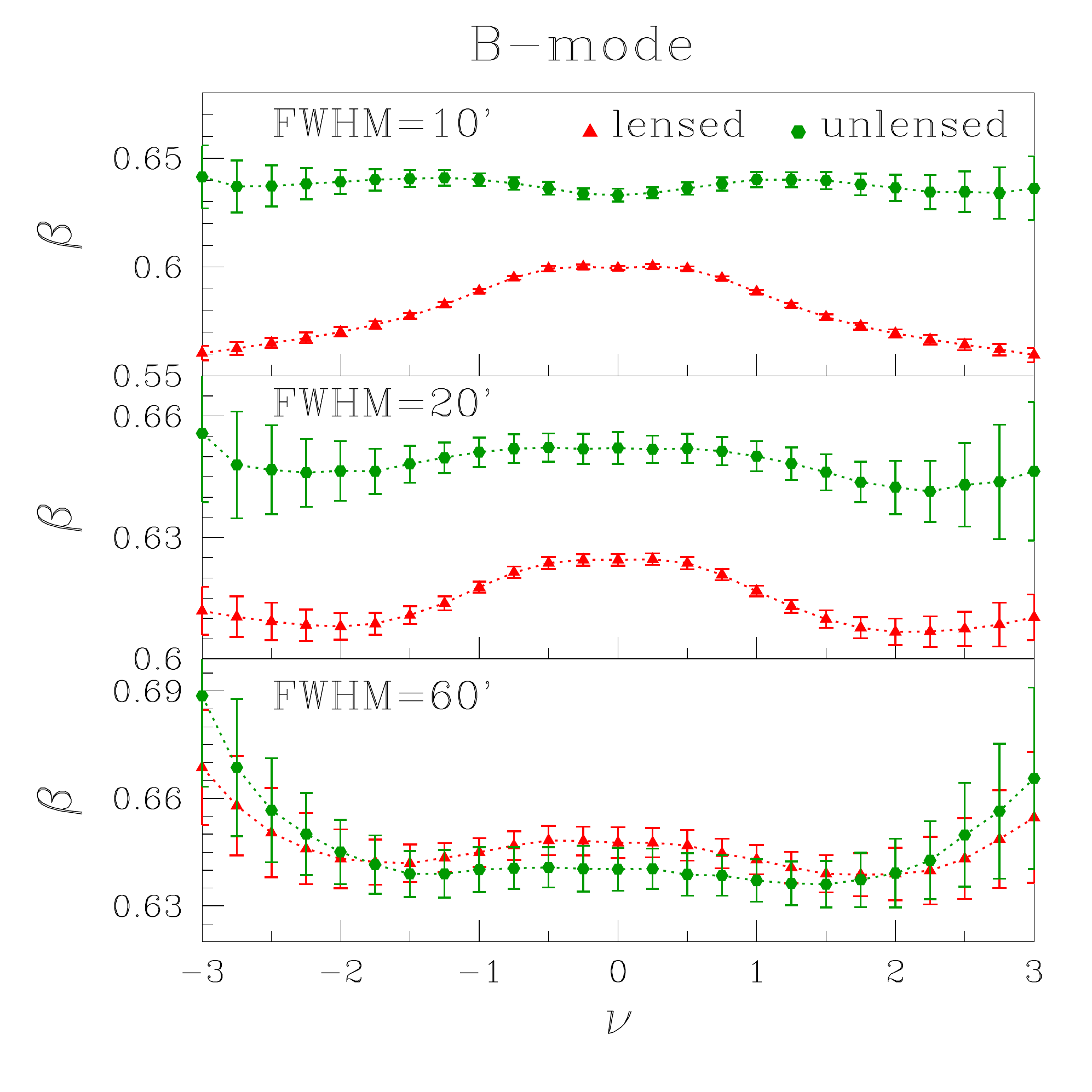}\\
\includegraphics[height=2.in,width=2in]{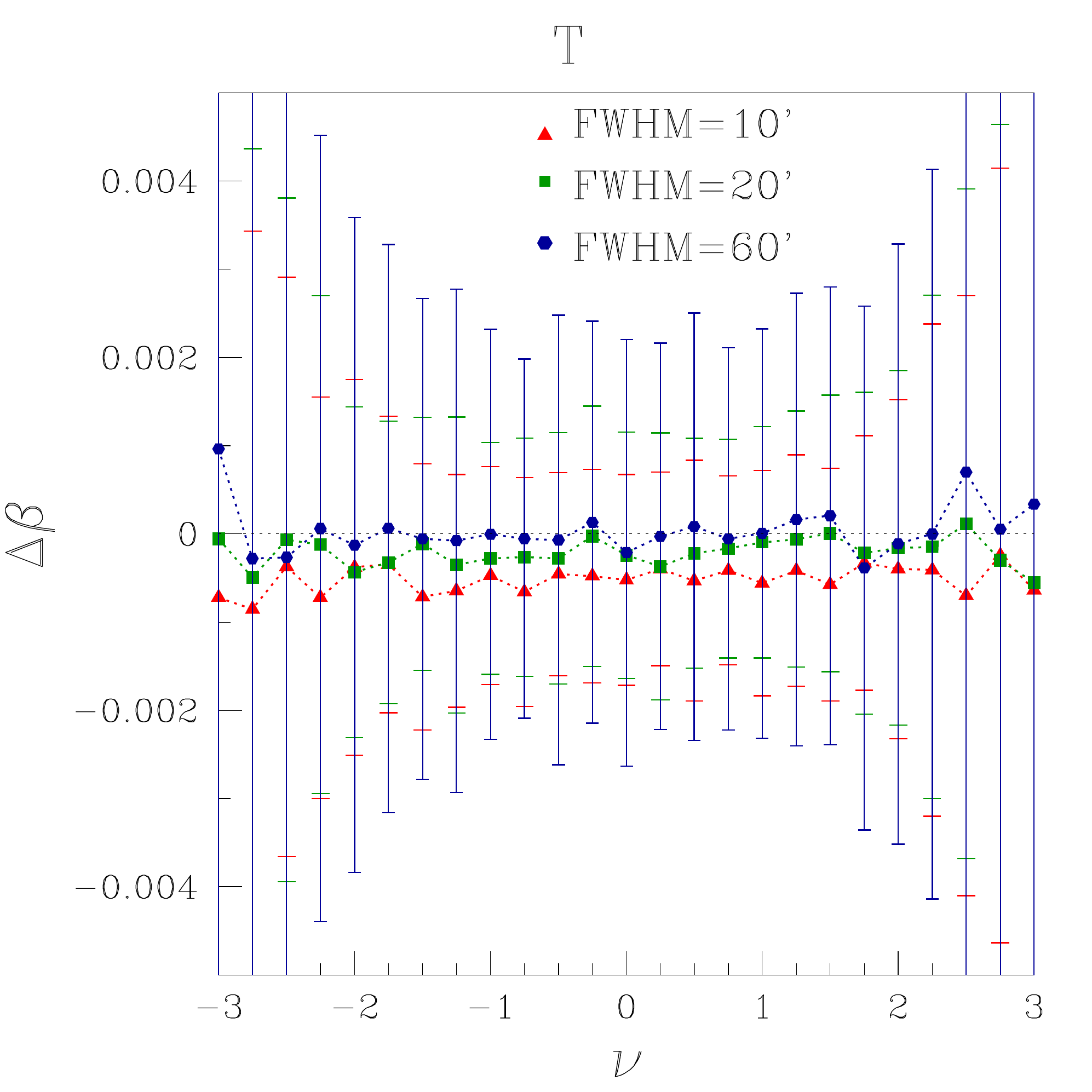}
\includegraphics[height=2.in,width=2in]{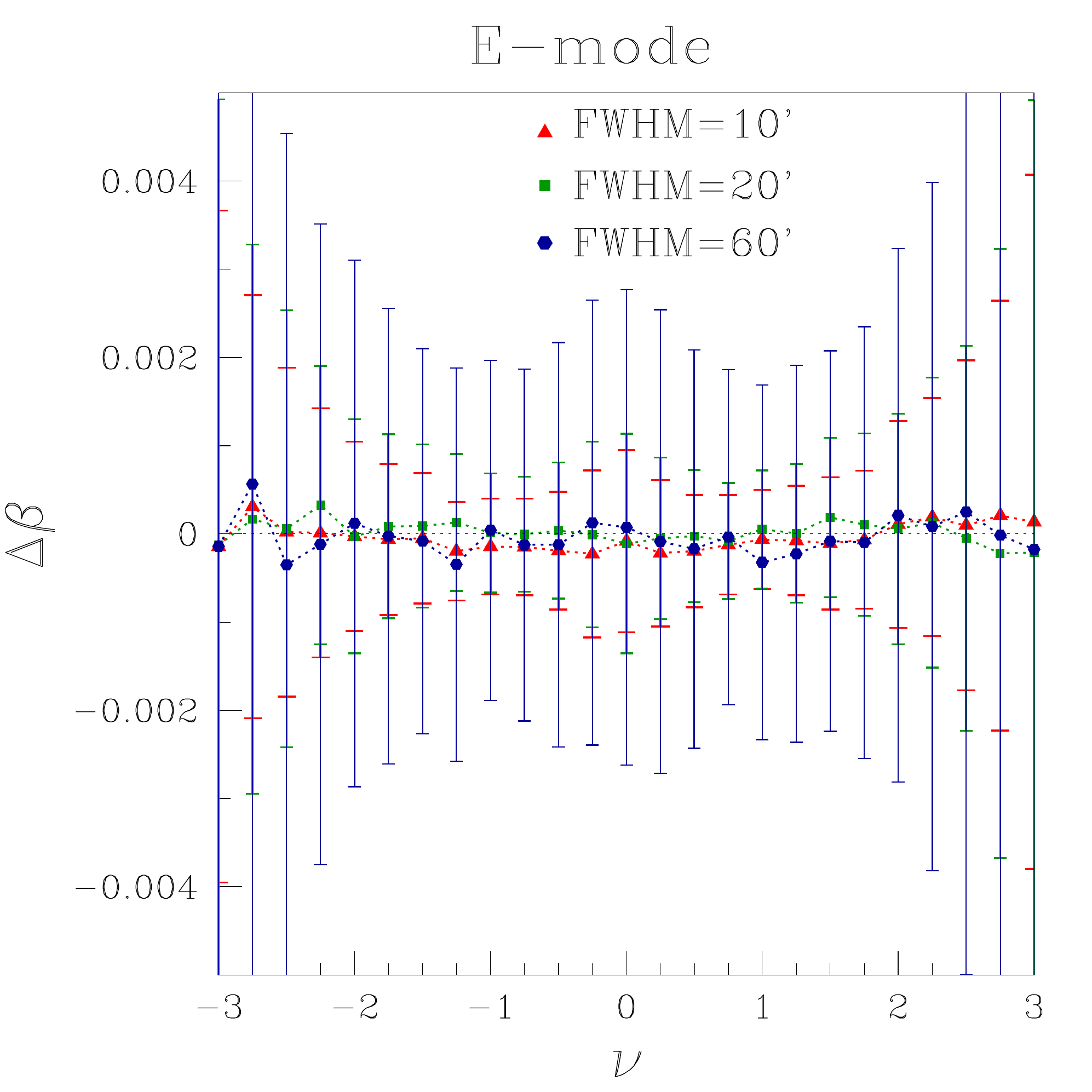}
\includegraphics[height=2.in,width=2in]{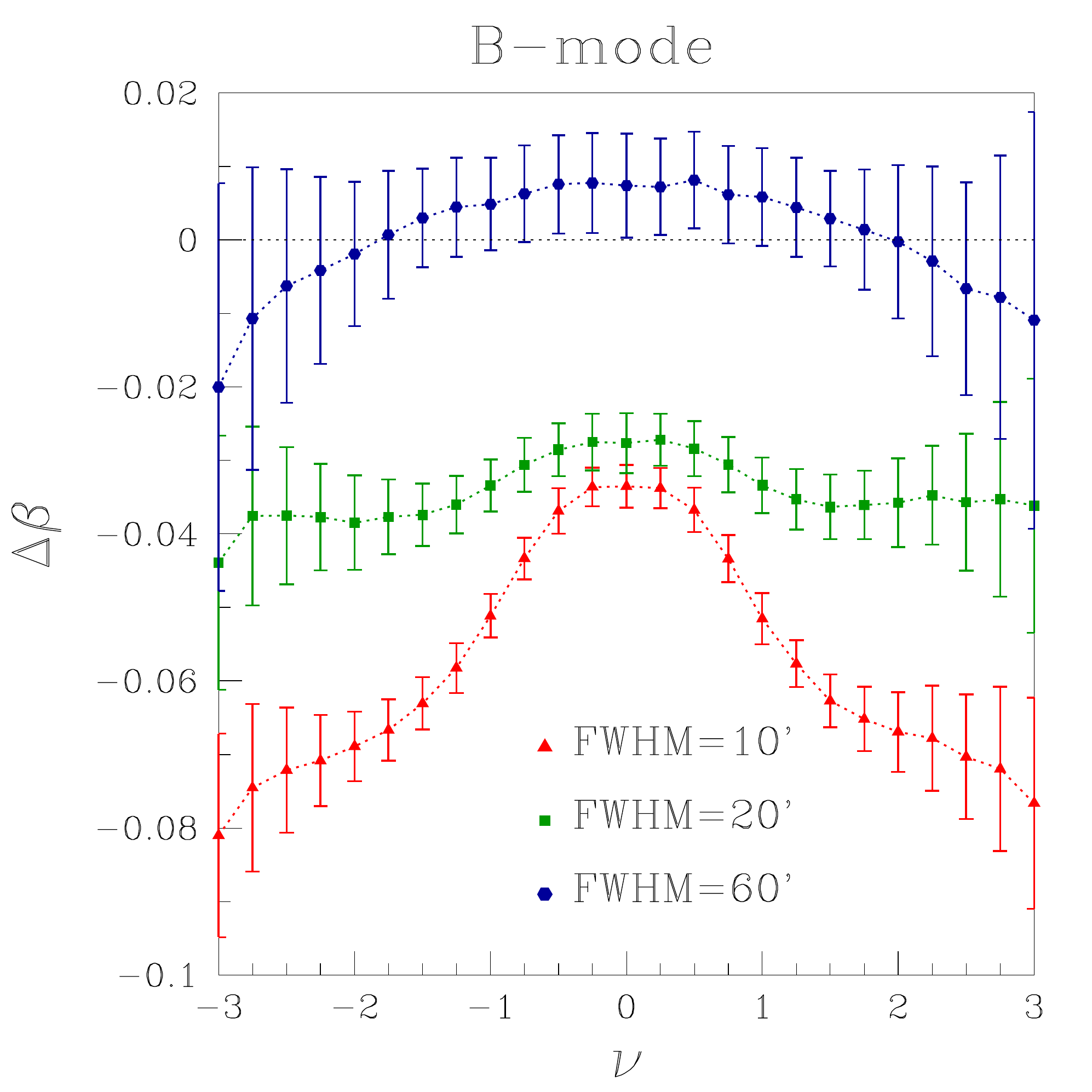}
\caption{{\em Top panels}: Mean values of $\beta$ for $T$ (left), $E$ (middle) and $B$ (right) for FWHM=20' for unlensed (green) and unlensed (red) cases. {\em Bottom panels}: The difference between lensed and unlensed mean $\beta$ values for the three fields shown in the top panels. All plots are average over 200 maps and error bars are the standard deviations obtained from them.}
\label{fig:beta}
\end{figure}
\end{center}

To quantify the deviation in the average $\beta$ induced by lensing let us denote
\begin{equation}
  \Delta \beta(\nu) \equiv   \beta^{\rm L}(\nu) - \beta^{\rm UL}(\nu).
\end{equation} 
$\Delta \beta(\nu)$ encapsulates the strength of the shear caused by lensing at different threshold levels. 
In the top panels of figure~\ref{fig:beta} we show $\beta$  versus $\nu$ for the lensed (red) and unlensed (green) maps,  for $T$ (left), $E$ (middle) and $B$ (right). The smoothing scales are the same as those in figure~\ref{fig:pdf_of_beta}, namely,  FWHM$=10',\ 20', \ 60'$. We can see that $\beta$ is symmetric about $\nu=0$ for both lensed and unlensed cases. The shape and amplitude of the curves  vary significantly for all the three fields.
Since we are dealing with smooth fields, it is reasonable to expect that at high values of $|\nu|$, the iso-field curves will follow the {\em curve shortening flow}\footnote{The nature of how the curves shorten with increasing $|\nu|$ and finding the evolution equation that governs it is an interesting mathematical question in its own right. For example for Gaussian fields we know the perimeter decreases exponentially with increasing $|\nu|$. We will, however, not address it here and pursue it elsewhere.}, under which the perimeter of the curves decrease and  converge to circular shape, and finally  collapses to a single point of singularity at local maxima and minima. Therefore, $\beta$ must tend towards one at sufficiently large $\nu$. The range of $\nu$ in figure~\ref{fig:beta} is chosen to highlight comparison across different smoothing scales. Note that the rms of the fields increase with decrease of smoothing scale. For FWHM=10', $\beta$ has turning points beyond $|\nu|\sim 3$ and does tend to one at large $|\nu|$.

The lensing distortions for $T$ and $E$ are not noticeable visually, while for $B$ the effect is very strong. 
The bottom panels of figure~\ref{fig:beta} show $\Delta\beta$ corresponding to the respective top panels. We can discern a weak but distinct negative value of $\Delta\beta$ for $T$ that increases with decrease of the smoothing scale. This implies that the lensing remapping causes structures to become statistically  more anisotropic at all threshold levels. The effect becomes more pronounced as we probe down to smaller scales. For $E$ mode, the deviations remain consistent with zero.

$\Delta\beta$ for $B$ mode are large and exhibits wide variation with smoothing scale. The values are increasingly negative at smaller smoothing scales, where the lensing effect is strongest. At the relatively larger smoothing of $1^{\circ}$ where we expect both primordial and lensed components to contribute to $\beta$ we find  that $\Delta\beta$ has is positive for a symmetric range around $\nu=0$, indicating that structures at these threshold range are more isotropic.

A comment regarding the impact of numerical errors arising from stereographic projection on the interpretation of our results for $\beta$ is in order at this point. Our results suggest that the effect of lensing on $T$ and $E$ that is encapsulated in $\beta$ is negligibly small at the smoothing scales that we have probed. For accurate calculation of the effect we stress that we should carry out the calculation directly on the sphere.  
For $B$ the distortion effects are large and increase with decrease of the smoothing scale. Since these are structures that arise due to lensing we conclude that this is a real signature of lensing induced distortions. However, the importance of calculating $\beta$ directly on the sphere cannot be overstated for precise quantification of lensing distortions.

\section{Conclusion and discussion}
\label{sec:sec5}

The geometrical and topological properties of cosmological random fields, are a vast source of physical information that have not yet been fully exploited. The applicability of analysis methods that rely on such pixel space properties depend on availability of observational data which are high resolution, cover large field of view, and have good signal to noise ratio. Rapid advances in observational technology  and imaging are making such good quality data increasingly more available across a wide  spectrum of length scales. Therefore, real pixel based analysis methods can be expected to become more relevant and useful in the near future. This paper is the latest in a series of investigations that extend previously used pixel space methods of studying cosmological fields. Our focus in this work is to understand morphological changes induced in the fields of the CMB by gravitational lensing due to intervening matter distribution. 

For our analysis, we have used shape parameters, $\alpha$ and $\beta$, defined from the eigenvalues of the CMT, to capture the information about the statistical isotropy and the net anisotropy of distribution of structures of a field. We have used two methods of calculation, the first one computes the CMT and hence, $\alpha$, directly on the sphere. The second method which is used to calculate $\beta$ is carried out on the flat plane after stereographic projection of the fields. Before probing the effect of lensing on CMB fields, we have studied the effect of varying cosmology and hemispherical anisotropy on $\alpha$ and $\beta$. The purpose of this exercise is to distinguish the effect of lensing from other physical phenomenon that can mask its effects. We find $\alpha$ to be insensitive to variation of cosmological parameters but it can capture the hemispherical anisotropy in CMB temperature field especially at large angular scales. $\beta$ for $T$, $E$ and $B$ fields shows very distinct dependence on the cosmological models as well as hemispherical anisotropy. However, the cosmological parameter values that we have considered are far off from the best fit values obtained from Planck data, and it is not clear whether the distinct signatures in $\beta$ can be used for constraining cosmology. 

Minkowski tensors, and in particular the CMT provides a novel approach to quantify the effect of lensing on the CMB. We have carried out a detailed investigation of the distortion effects induced by lensing that is encapsulated in $\alpha$ and $\beta$ for all the CMB fields. We do not find signatures of lensing on $\alpha$ in the range of scales that we have studied for temperature and $E$ mode. For $B$ mode, its coupling with $E$ mode induced by lensing at smaller angular scales is manifest as positive values of the difference of $\alpha$ between lensed and unlensed maps. We find that this difference peaks at around FWHM=5' corresponding to smoothing angle of 2.12'.
Further, we find that statistical isotropy is preserved for the lensed fields as is expected from isotropic large scale matter distribution in the Universe.

Next we have quantified the distortions induced by lensing encapsulated by $\beta$. Our results imply that lensing makes structures more anisotropic across all threshold levels for CMB temperature. We find that the level of distortion increases towards smaller angular scales in the range of scales that we have studied. For $E$ mode we do not obtain a clear trend for the distortion effects. $B$ mode on the other hand exhibits very strong effect with increasing anisotropy at smaller scales. Our analysis here for $\beta$ for all the fields is limited to relatively large angular scales ($\ge 10'$) for two reasons - use of stereographic projections and limited computing resources. It will be useful to probe down to smaller scales of around 2' as done for $\alpha$.   

The work presented in this paper can be extended in the following directions. First, our results are based on pure simulations. For applications to observed data we need to take into account the effects of instrumental noise and residual foregrounds. Secondly, it is desirable that the computation of $\beta$ is done directly on the sphere without  projection to the plane for accurate estimation. We are developing a numerical code for this purpose. Thirdly, for a comprehensive analysis of the lensing effects we should use the full suite of scalar and tensorial Minkowski functionals and Betti numbers. Our approach differs from previous use of scalar MFs in that we compute the geometry (namely, area, perimeter and counts) of individual structures of the excursion sets and look at their statistics, rather than simply their mean values. However, such calculations must be carried out directly on the sphere in order to avoid projection effects. We plan to use the results obtained here as a guide to deeper understanding of the lensing deflection field or lensing potential field.

\acknowledgments
{We acknowledge the use of the \texttt{Hydra} and \texttt{NOVA} HPC clusters at the Indian Institute of Astrophysics. Some of the results in this paper have been obtained by using the \texttt{CAMB}~\cite{Lewis:2000ah,cambsite},  \texttt{HEALPIX}~\cite{Gorski:2005,Healpix} and \texttt{LENSPIX}~\cite{Lenspix} packages. The work of P.~Chingangbam is supported by the Science and Engineering Research Board of the Department of Science and Technology, India, under the \texttt{MATRICS} scheme, bearing project reference no \texttt{MTR/2018/000896}.}

\appendix
\section{Effect of stereographic projection on the calculation of $\beta$}
\label{sec:appendA}

Stereographic projection of structures from the two-sphere to the plane preserves the shape, but not area. As such, we naively expect that the statistic $\beta$ will remain unaffected by such a mapping. However, there are subtleties associated with this statement. 

To extract $\beta$ from a set of structures on the plane, we utilise a numerical algorithm that generates iso-field lines obtained by linearly interpolating between field values on a discretized, regular grid. A numerical error will be associated with this procedure, due to the discretization of a smooth field. This error will be a function of the size of a structure -- the more pixels that it encompasses, the better the structure is resolved and the more accurately we can determine its shape. 

In terms of projecting from the sphere to the plane, the result is a systematic error in our estimation of the shape of structures.  The size of an object after projection will monotonically decrease if its location on the sphere varies with polar angle from $\theta=0$ (North pole, appears largest) to $\theta = \pi$ (south pole, appears smallest).

We can estimate the magnitude of this effect by projecting a known structure from the sphere to the plane, and calculating $\beta$. We generate a disc of radius $r = \pi/4\ {\rm rad}$ on the two-sphere, with centre randomly located over the range $0 < \phi < 2\pi$ and $\pi/2 < \theta < \pi$, where $\theta, \phi$ are the polar and azimuthal angles respectively. We then stereographically project this structure onto a regular two-dimensional grid, and define a binary field on the plane as $\delta_{ij} = 1$ if one of the disc pixels on the sphere is projected onto the plane pixel $i,j$, and $\delta_{ij} = 0$ otherwise. The disc, after projected onto the plane, in principle retains its shape\footnote{subject to it not intersecting with the point $\theta = 0$} however its size will vary as the centre of the disc moves from $\theta=\pi$ (smallest) to $\theta = \pi/2$ (largest) on the two-sphere. As a result, its $\beta$ parameter, which should satisfy $\beta = 1$, will be reduced and be a function of $\theta$; $\beta = \beta(\theta) < 1$.

To test the magnitude of this systematic, we randomly place a disc of radius $r = \pi/4$ radians on the sphere $N_{\rm real} = 5000$ times and repeat the test. The center of the disc is selected to be within the range $0< \phi < 2\pi$ and $\pi/2 < \theta < \pi$. We repeat using four combinations of grid sizes -- we take two, equal area Healpix  grids on the sphere with number of pixels $n_{\rm side} = 256$ and $n_{\rm side} = 512$ and create a regular grid on the plane with pixels $n_{\rm pl} = 512^{2}$ and $n_{\rm pl}=1024^{2}$. The plane size is fixed as $-5 < x,y < 5$, sufficiently large to encompass all projections. From each binary $\delta_{ij}$ field, we draw an iso-field line at $\delta = 0.5$, and calculate $\beta$ for this shape. For a perfectly resolved projected circle, we can expect $\beta=1$.

\begin{figure*}
  \includegraphics[width=0.48\textwidth]{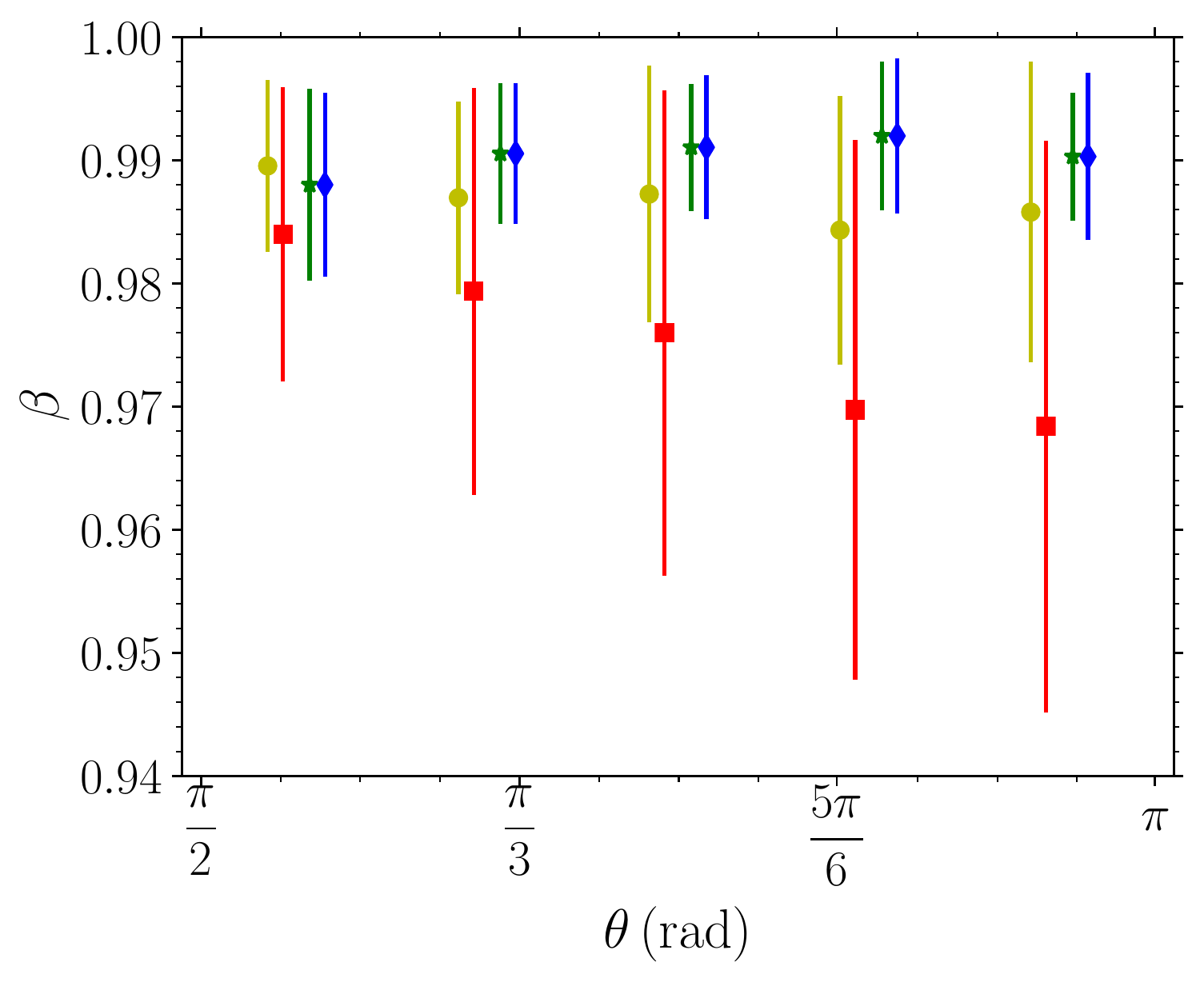}
    \includegraphics[width=0.48\textwidth]{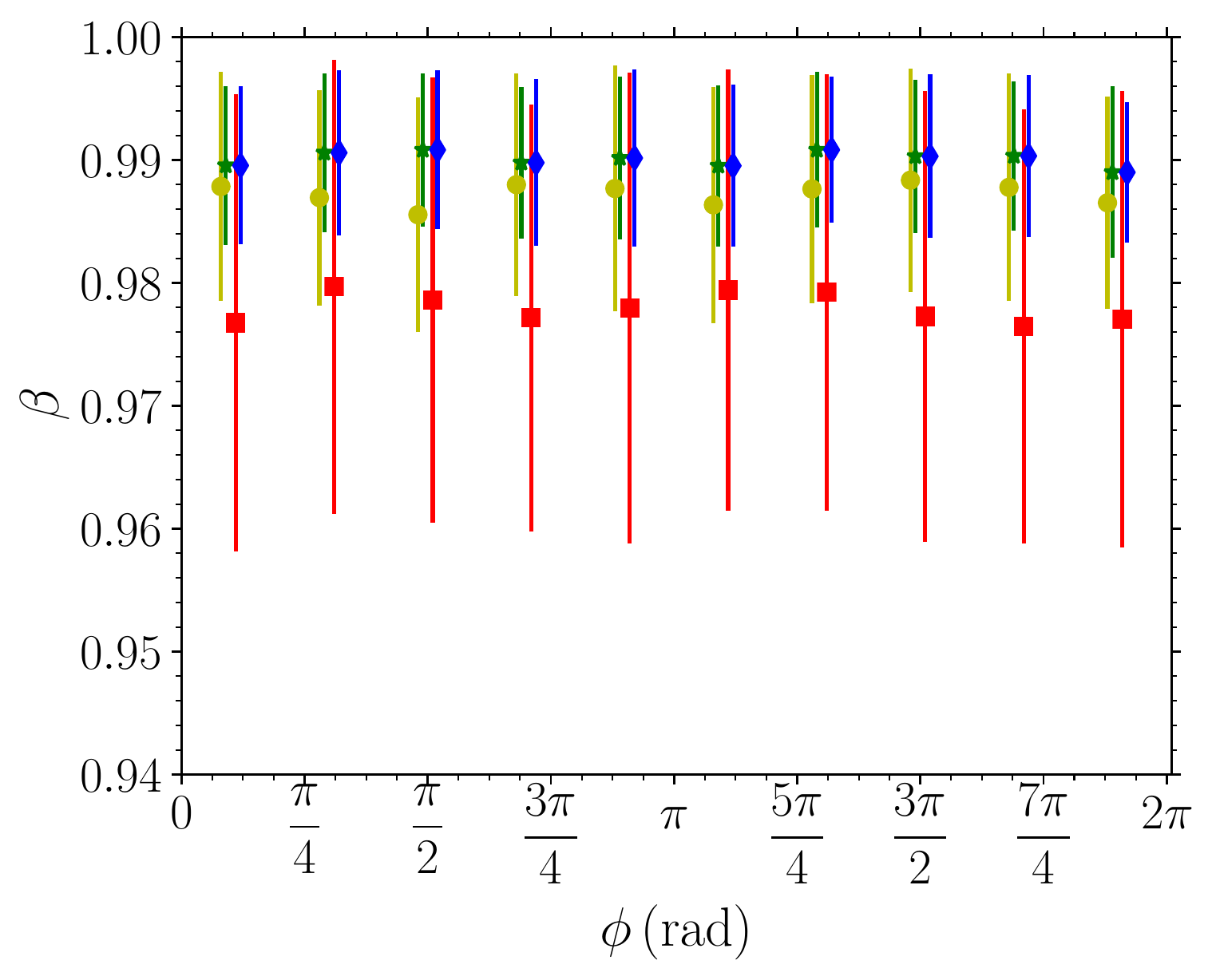}
  \caption{{\em Left panel:} $\beta$ as a function of polar angle $\theta$. The yellow/red/green/blue points correspond to $(n_{\rm side}, n_{\rm pl}) = (512,512^{2}),(256,512^{2}),(512,1024^{2}),(256,1024^{2})$ respectively. {\em Right panel:} $\beta$ as a function of azimuthal angle $\phi$. We observe no systematic dependence on the shape of the projected disc and $\phi$, as expected.}
  \label{fig:beta_theta_phi}
\end{figure*}

In the left panel of Figure~\ref{fig:beta_theta_phi} we present $\beta$ as a function of $\theta$.  The points and error bars are the sample mean and standard deviation of the measurements. The yellow/red/green/blue points correspond to $(n_{\rm side}, n_{\rm pl}) = (512,512^{2}),(256,512^{2}),(512,1024^{2}),(256,1024^{2})$ respectively. One can observe a systematic decrease in $\beta$ with the position of the disc on the sphere for the most poorly resolved system $(n_{\rm side}, n_{\rm pl}) = (256,512^{2})$ (red points). As expected, near the equator at $\theta = \pi/2$, the shape of the object is best reconstructed $\beta \to 1$. Objects nearest the south pole of the sphere are most poorly reconstructed and have a marginally lower $\beta$ value. However, for higher resolutions of the plane and sphere, this systematic behaviour becomes negligible.

In the right panel of Figure \ref{fig:beta_theta_phi} we present the $\beta$ values as a function of azimuthal angle $\psi$. We observe no correlation between $\beta$ and $\phi$ as expected.

\section{Sensitivity of $\alpha$ and $\beta$ to variation of cosmological parameters}
\label{sec:alpha_vary_lcdm}

In order to test the sensitivity of $\alpha$ and $\beta$ to cosmology, as discussed in section~\ref{sec:alpha_lcdm}, we consider three different sets of cosmological parameters, in addition to the {\em fiducial} $\Lambda$CDM model. The value of the parameters for the models considered are given in Table~\ref{table:models}.  Model 1 has 95\% cold dark matter and 5\% baryons, keeping $n_s$ the same as the fiducial model. Models 2 and 3 have different values of $n_s$ while the matter density fractions are the same as the fiducial model. We assume flat universe for all the models under consideration. We have chosen models with unrealistic parameters values far off from the fiducial model so as to exaggerate the effect on $\alpha$ and $\beta$.

\begin{center}
\begin{figure}
\includegraphics[height=2.in,width=2.3in]{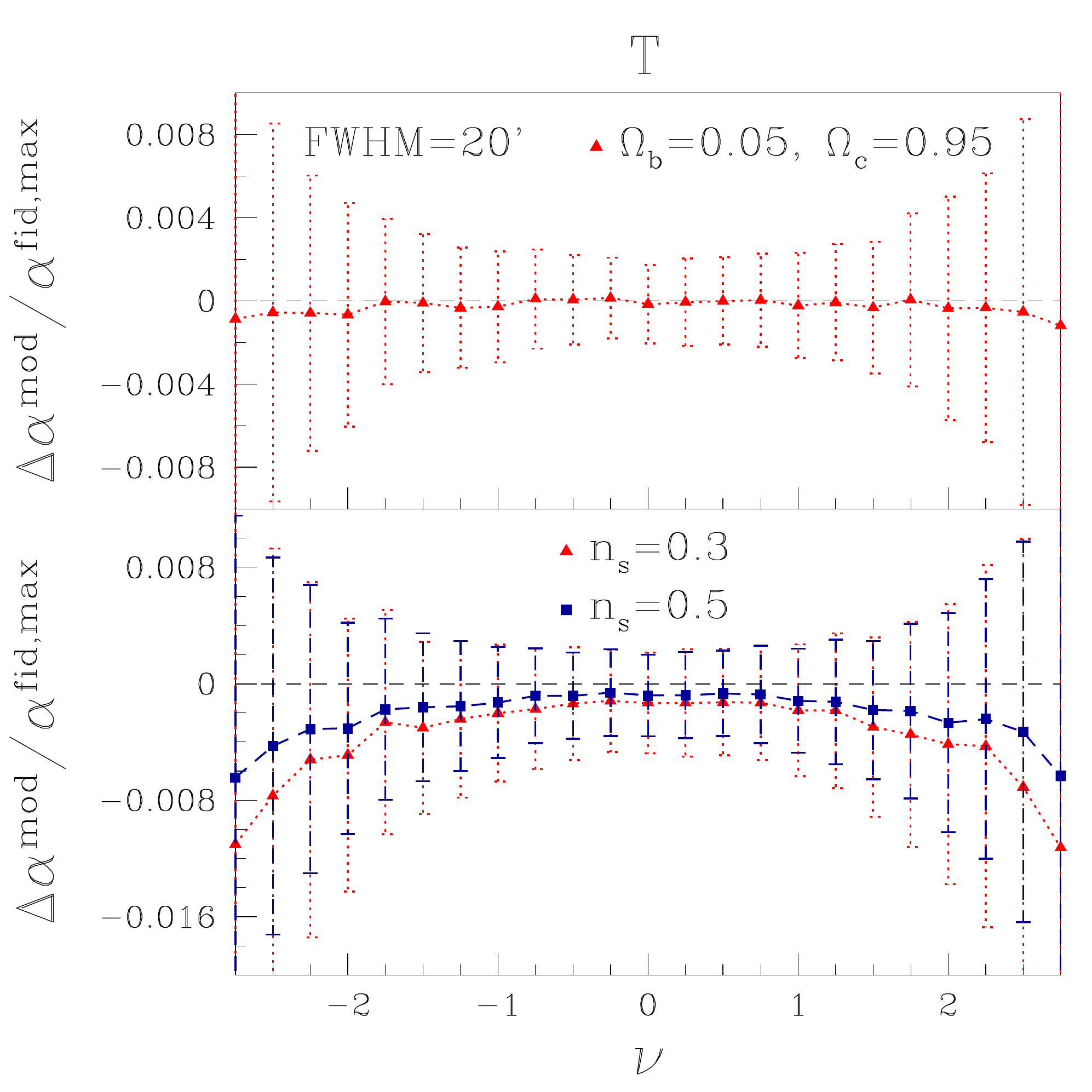}\quad\quad
\includegraphics[height=2.in,width=2.3in]{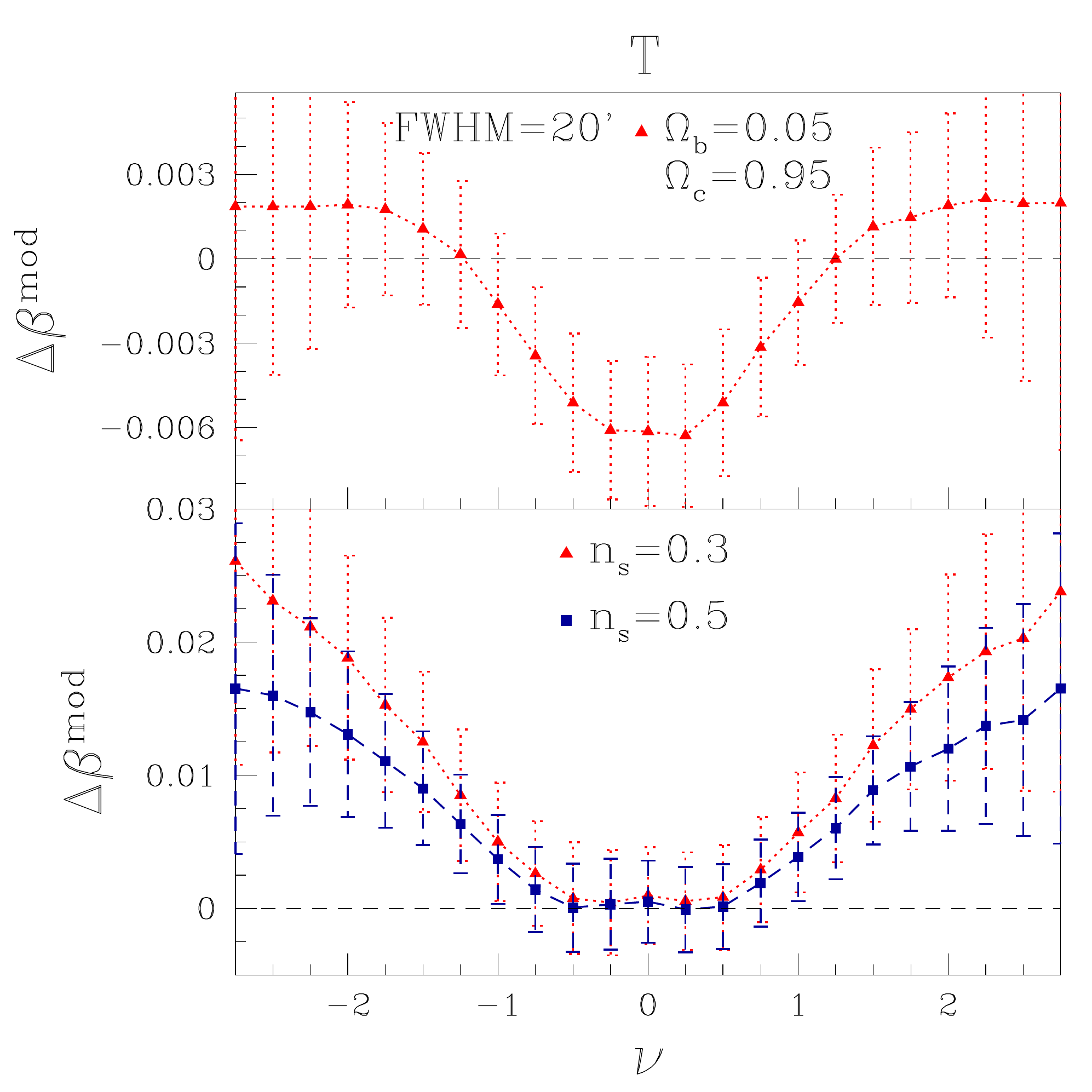} \\ 
\includegraphics[height=2.in,width=2.3in]{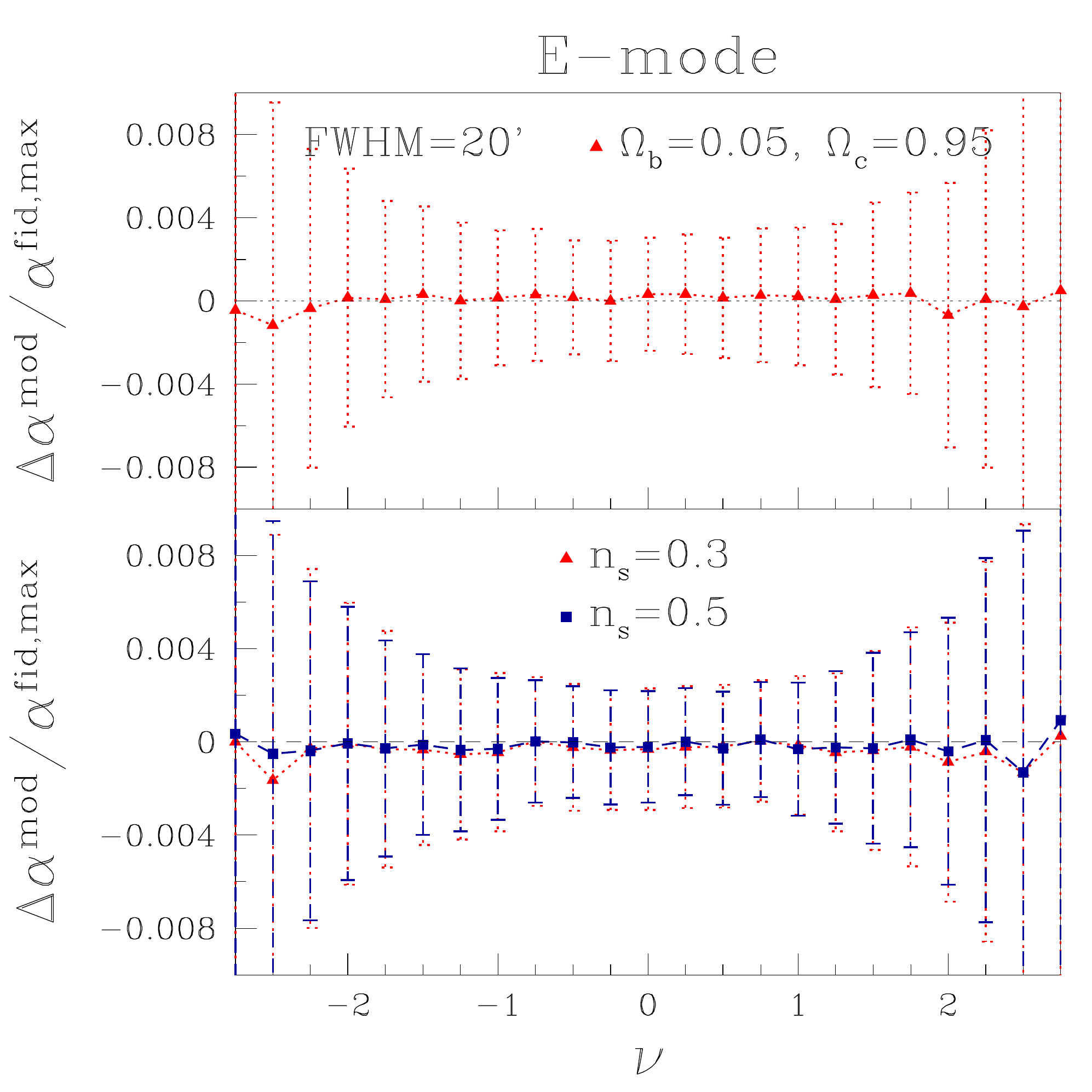}\quad\quad
\includegraphics[height=2.in,width=2.3in]{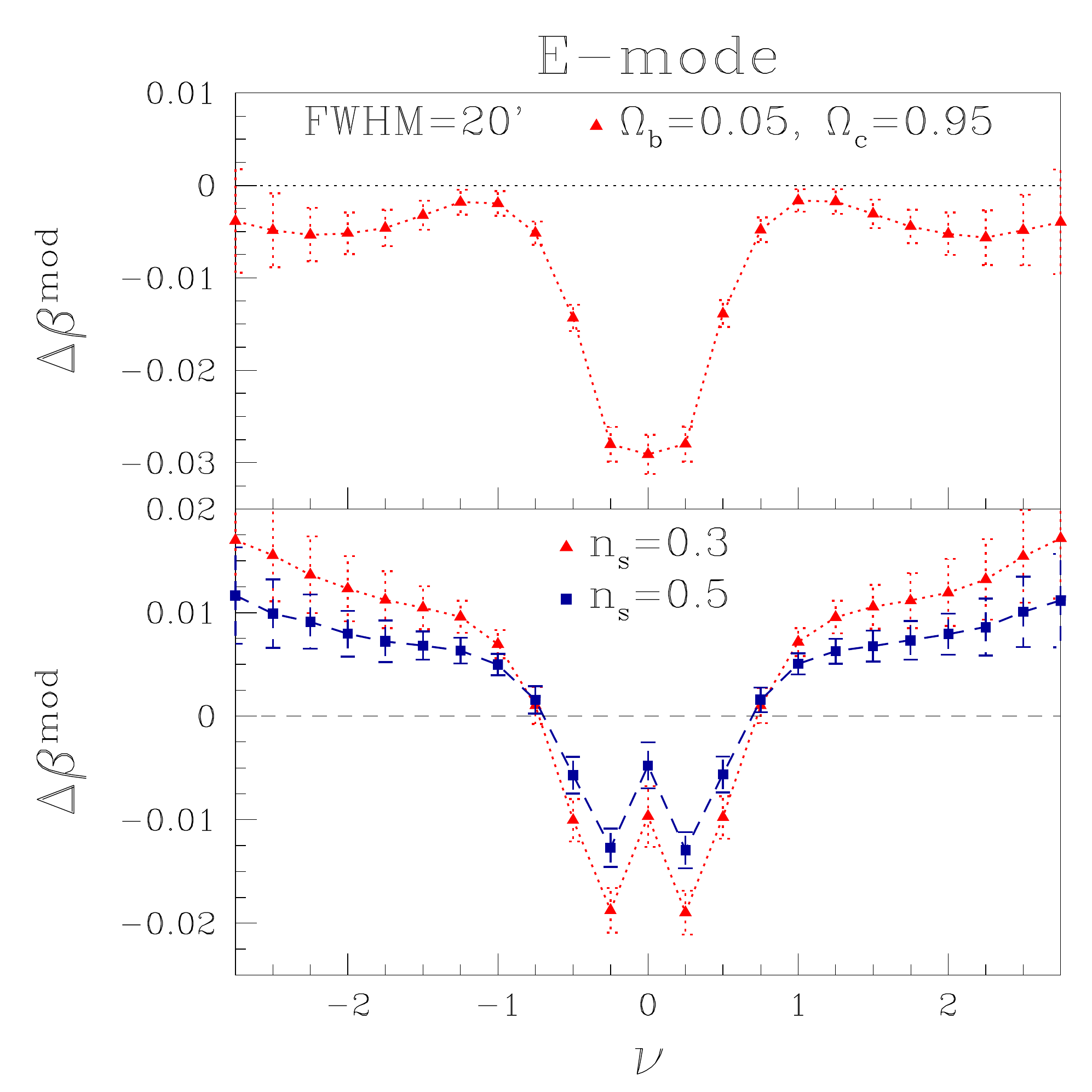}\\
\includegraphics[height=1.6in,width=2.3in]{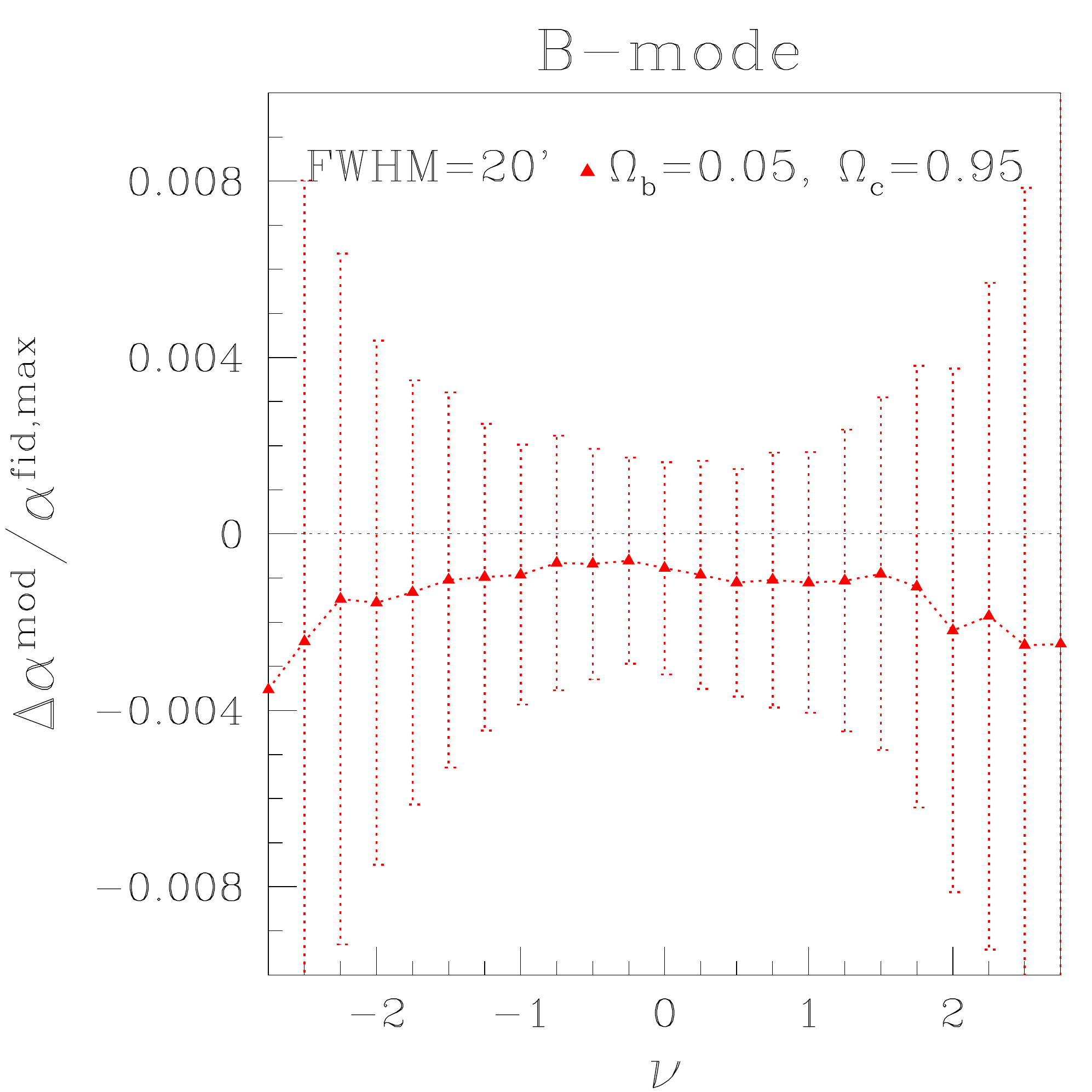}\quad\quad
\includegraphics[height=1.6in,width=2.3in]{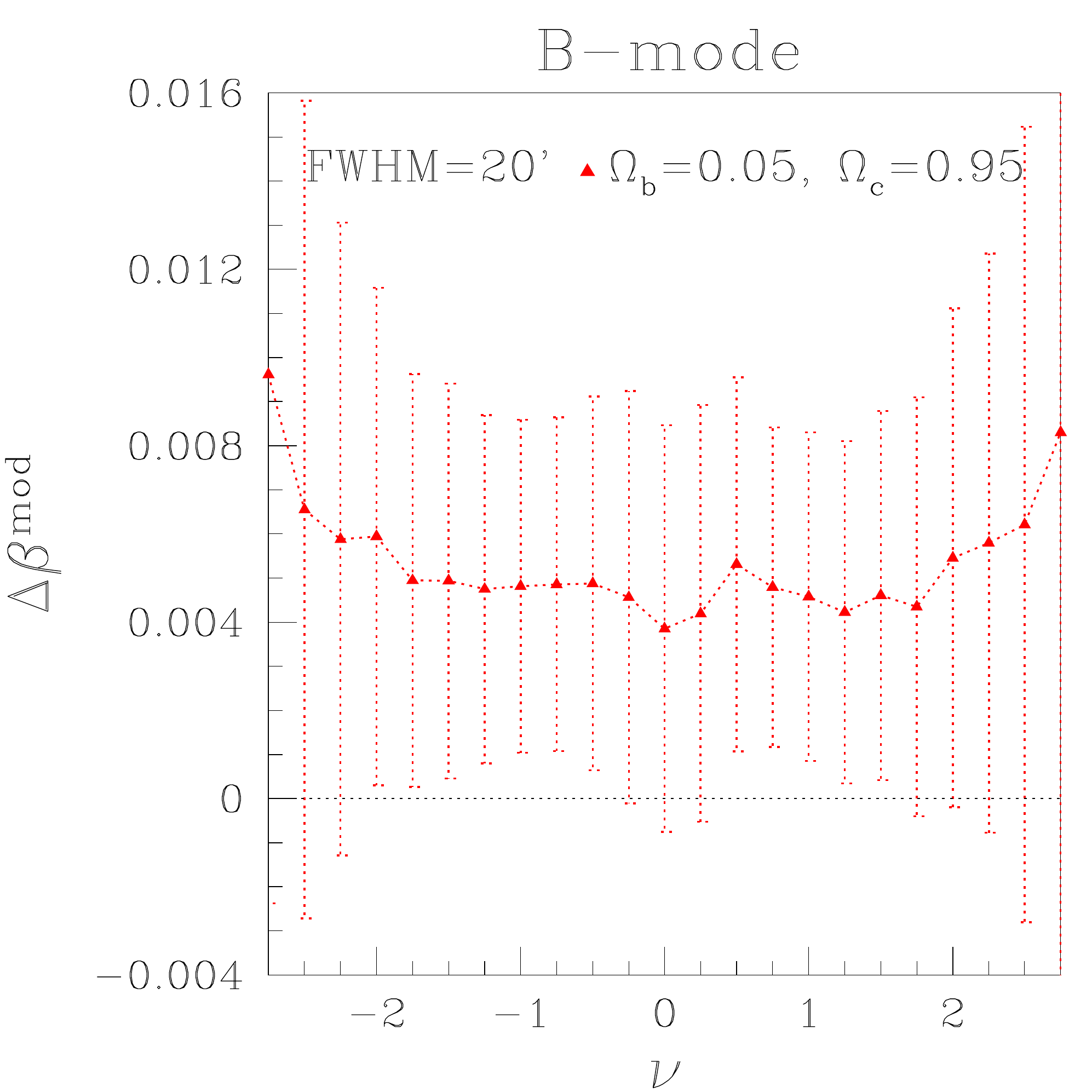}
\caption{Deviations of $\alpha$ (left column) and $\beta$ (right column) for different cosmological models (listed in Table~\ref{table:models}) from the `fiducial' $\Lambda$CDM model, for $T$ (top), $E$ (middle) and $B$ mode (bottom)  fields. The smoothing scale used is FWHM=20'. The upper panels in each plot show the effect of varying the matter density fraction, while the lower panels show the effect of varying $n_s$. For $B$-mode there is only one panel since we have fixed the tensor spectral index to be one. We have assumed a flat universe for all the cases.}
\label{fig:alpha_beta_om}
\end{figure}
\end{center}

Let us denote the difference of $\alpha$ between each non-standard model (superscript `mod') and fiducial model (superscript `fid') by
\begin{equation}
  \Delta\alpha^{\rm mod} \equiv \alpha^{\rm mod} - \alpha^{\rm fid}.
\end{equation}
Then we normalize this quantity by the value of $\alpha^{\rm fid}$ at $\nu=0$. 
The left column of  figure~\ref{fig:alpha_beta_om} shows plots for normalized $\Delta\alpha^{\rm mod}$ for $T$ and $E$ and $B$ modes. For $T$ and $E$ upper panels correspond to variation of matter content while lower panels correspond to variation of $n_s$. For $B$ we only vary the matter content and fix the tensor spectral index to be one. All plots are average over 500 realizations and the error bars are the corresponding sample variance. The $y$-axis scales are the same for all plots.

\begin{table}
\begin{center}
  \begin{tabular}{|lllll|}
  \hline
  Fiducial model & $\Omega_{\rm cdm}= 0.228$, &  $\Omega_{\rm b}= 0.046$,  & $\Omega_{\Lambda}= 0.726$, &  $n_s=0.960$ \\
  \hline
     Model 1 & $\Omega_{\rm cdm}=0.950$, &  $\Omega_{\rm b}=0.050, $  & $\Omega_{\Lambda}= 0.0, $ &  $n_s=0.960$ \\
      \hline
     Model 2 & $\Omega_{\rm cdm}= 0.228$, &  $\Omega_{\rm b}=0.046, $  & $\Omega_{\Lambda}= 0.726, $ &  $n_s=0.50$ \\
        \hline
      Model 3 & $\Omega_{\rm cdm}=0.228$ &  $\Omega_{\rm b}=0.046$,  & $\Omega_{\Lambda}= 0.726$, &  $n_s=0.30$ \\
  \hline
  \end{tabular}
\end{center}
  \caption{Models with different values of cosmological parameters for which we compute $\alpha$. The input for the amplitude of primordial power spectrum and reionization history is the same in all the models. We assume curvature parameter, $\Omega_k$, is zero for all models. All models other then the fiducial one have unrealistic parameter values so as to magnify their effect in $\alpha$ and $\beta$.}
\label{table:models}
\end{table}

{\em Effect of varying matter density fractions on $\alpha$:} We find that $\Delta\alpha^{\rm mod}$ for  $T$ and  $E$ are negligible, in agreement with our expectation. For $B$, $\Delta\alpha^{\rm mod}$ is small but negative. This can be explained by the fact that the absence of dark energy decreases the luminosity distance to the last scattering surface, which results in fewer number of structures per unit solid angle. Since the same reasoning applies to $T$ and $E$ also, similar negative value should be obtained. However, due to the relatively larger number structures of these two fields the effect is not pronounced.  For realistic range of matter fractions within PLANCK constraints, we can safely assume that $\alpha$ is insensitive to cosmology.

{\em Effect of varying $n_s$ on $\alpha$:} Lower values of $n_s$ leads to reduced power in the primordial power spectrum for modes $k>k_0$, where $k_0$ is the pivot scale which is chosen by \texttt{CAMB} to be $0.05\, {\rm Mpc}^{-1}$. This implies fewer number of structures on small length scales.  This is the reason why we obtain $\Delta\alpha^{\rm mod}$ negative, shown in the lower panels of the plots for $T$ and $E$ mode where smoothing scale FWHM=20' is sufficiently small scale. Again, as in the case of variation of matter fractions, for realistic range of $n_s$ within PLANCK constraints, the variation on $\alpha$ is very small, and we can safely ignore it.

Next we discuss the effect of cosmology on $\beta$. Let us denote
\begin{equation}
  \Delta\beta^{\rm mod} \equiv \beta^{\rm mod} - \beta^{\rm fid},
\end{equation}
where the superscripts have the same meaning as in the case of $\alpha$.
The right column of  figure~\ref{fig:alpha_beta_om} shows plots for $\Delta\beta^{\rm mod}$ for $T$ and $E$ and $B$ modes. For $T$ and $E$ upper panels correspond to variation of matter content while lower panels correspond to variation of $n_s$. Again, for $B$ we only vary the matter content and fix the tensor spectral index to be one. All plots are average over 300 realizations and the error bars are the corresponding sample variance. Note that we have used different $y$-axis scales for the different models for the three fields. We find that $\Delta\beta^{\rm mod}$ has distinct shapes as the function of $\nu$ for the different fields. For understanding this behaviour it is useful to remind ourselves that structures at different $\nu$ correspond to different scales of clumpiness or emptiness.  
The distinct shapes of $\Delta\beta^{\rm mod}$ informs us that the anisotropy at different $\nu$ is affected differently by variations in matter fractions. A comparison of the amplitude of deviation for the three fields shows that $E$-mode has the highest level of deviations. These findings can be potentially useful in inferring properties of the universe. 

\section{Sensitivity of $\alpha$ and $\beta$ to Hemispherical anisotropy}
\label{sec:alpha_ha}

\begin{center}
\begin{figure}[htb]
\includegraphics[height=2.5in,width=2.7in]{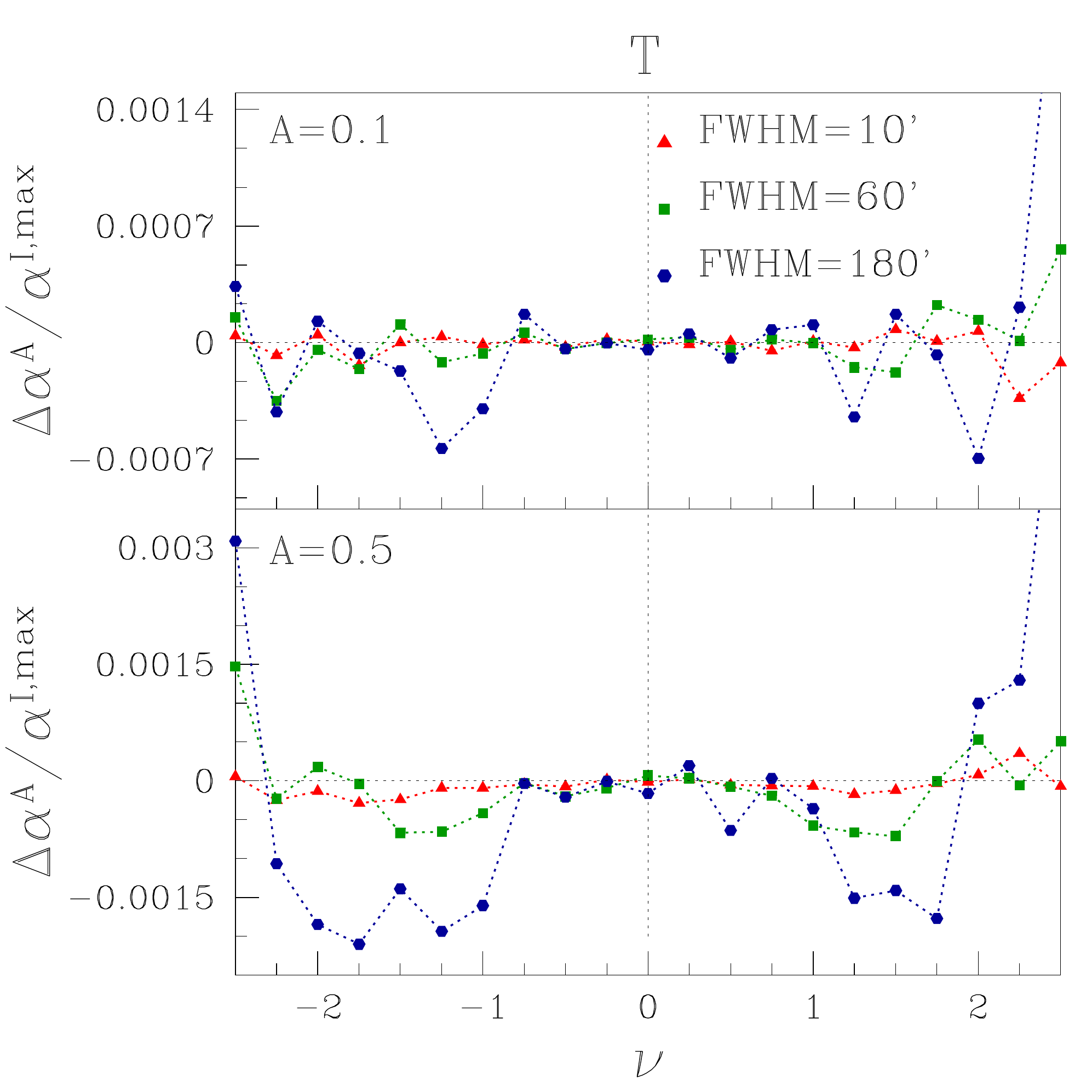}
\includegraphics[height=2.5in,width=2.7in]{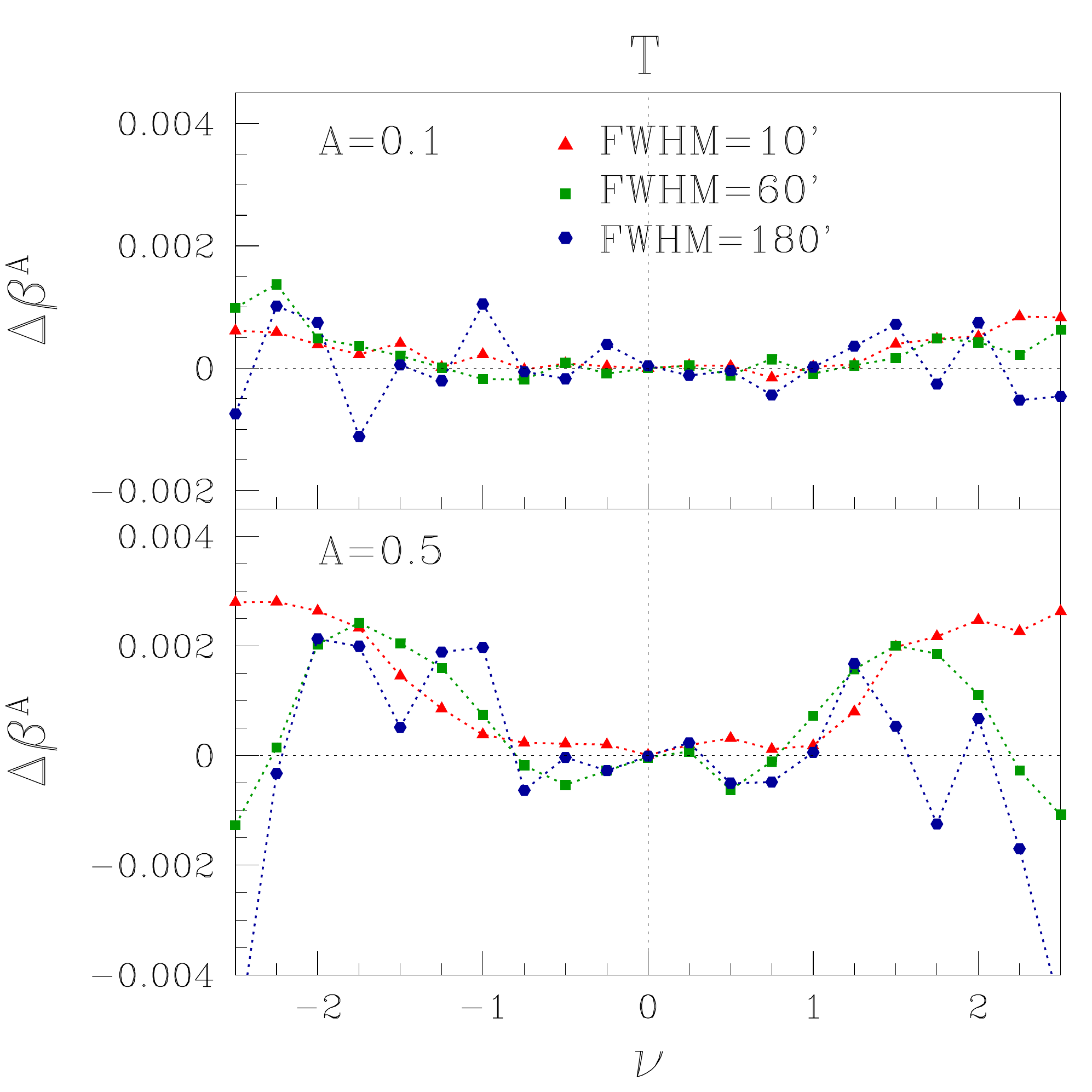}
\caption{{\em Left}: $\Delta\alpha^A$ for temperature field with input hemispherical anisotropy corresponding to $A=0.1$ and 0.5, for three different smoothing scales  at FWHM=10', 60' and 180'. All plots are average over 500 maps. We can see that the deviation in $\alpha$ becomes distinguishable as we increase the smoothing scale. {\em Right}: $\Delta\beta^A$ for the same fields, values of $A$, and smoothing scales as for $\alpha$.  We have not added error bars since they are roughly one order of magnitude larger than the mean values, and hence they make the mean values difficult to discern visually.}
\label{fig:alpha_aniso}
\end{figure} 
\end{center}

To study the effect of hemispherical anisotropy given in section~\ref{sec:aniso} we calculate the differences $\Delta\alpha^{\rm A} \equiv \alpha^{\rm A} - \alpha^{\rm I}$ and $\Delta\beta^{\rm A} \equiv \beta^{\rm A} - \beta^{\rm I}$, where the suffix A stands for the anisotropic case and I denotes the isotropic case.

In the left panel of figure~\ref{fig:alpha_aniso} we show $\Delta\alpha^{\rm A}$, normalized by the value of $\alpha^{\rm I}$ at $\nu=0$, for different smoothing scales, and for two values of the anisotropy strength, $A=0.1$ (upper panel) and $A=0.5$ (lower panel). The plots are average over 500 realizations of CMB maps. As expected,  $\Delta\alpha^{\rm A}$ is larger for greater value of anisotropy strength parameter i.e. $A$. We observe that the effect of hemispherical anisotropy gets washed out at high resolution (small smoothing scale). This is due to the fact that at high resolution the number of structures is very high.  The effect of the anisotropy becomes noticeable at lower resolution. It is interesting to note that $\Delta\alpha^{\rm A}$ remains symmetric and that the difference is most pronounced at roughly $|\nu|\sim 1$.

The right panel of figure~\ref{fig:alpha_aniso} shows $\Delta\beta^{\rm A}$, for the same smoothing scales and anisotropy strengths as for $\alpha$. The plots are average over 300 maps. Again as expected,  $\Delta\beta^{\rm A}$ is larger for larger $A$. We observe that $\Delta\beta^{\rm A}$ has roughly the same shape and amplitude across the threshold range $-2.5<\nu<2.5$ for all smoothing scales. The positive values of $\Delta\beta^{\rm A}$ across this threshold range indicates that the structures are more anisotropic as a consequence of the anisotropy term in eq.~\ref{eqn:dT_aniso}. The level of anisotropy is larger towards larger $|\nu|$ albeit with larger error bars. Moreover, the statistical fluctuation increases for larger FWHM because the number of structures and corresponding total perimeter decreases with increasing FWHM. It is again interesting to note that $\Delta\beta^{\rm A}$ is symmetric. 
Our findings in this subsection suggest that $\alpha$ and $\beta$ are complementary and together they can be useful probe of departures from statistical isotropy of the universe.



\end{document}